\documentclass{aastex62}

\usepackage{amsmath}
\usepackage{xcolor}
\usepackage{footmisc}
\usepackage{rotating,longtable}

\received{----, 2020}
\revised{----, 2020}
\accepted{----, 2020}
\submitjournal{ApJ}

	\shorttitle{NGC 6910}
	\shortauthors{Harmeen et al.}

\begin{document}

	\title{Unveiling the physical conditions in NGC 6910}

	\correspondingauthor{Harmeen Kaur}
	\email{harmeenkaur.kaur229@gmail.com}

	\author{Harmeen Kaur}
	\affil{Center of Advanced Study, Department of Physics \\
	DSB Campus, Kumaun University Nainital, 263002, India}
	\author{Saurabh Sharma}
	\affil{Aryabhatta Research Institute of Observational Sciences (ARIES),
	Manora Peak, Nainital 263 002, India}
	\author{Lokesh K. Dewangan}
	\affil{Physical Research Laboratory, Navrangpura, Ahmedabad - 380 009, India}
	\author{Devendra K. Ojha}
	\affil{Tata Institute of Fundamental Research (TIFR),
	Homi Bhabha Road, Colaba, Mumbai - 400 005, India}
	\author{Alok Durgapal}
	\affil{Center of Advanced Study, Department of Physics \\
	DSB Campus, Kumaun University Nainital, 263002, India}
	\author{Neelam Panwar}
	\affil{Aryabhatta Research Institute of Observational Sciences (ARIES),
	Manora Peak, Nainital 263 002, India}



	\begin{abstract}
	Deep and wide-field optical photometric observations along with 
	multiwavelength archival datasets
	have been employed to study the physical properties of the cluster
	NGC 6910. The study also examines the impact of massive stars to their
	environment. The age, distance and reddening of the cluster are estimated 
	to be $\sim$4.5 Myr, $1.72\pm0.08$ kpc, and $ E(B-V)_{min}= 0.95$ mag, 
	respectively. The mass function slope ($\Gamma = -0.74\pm0.15$ in the 
	cluster region is found to be flatter than the Salpeter 
	value (-1.35), indicating the presence of excess number of massive stars.
	The cluster also shows mass segregation towards the 
	central region due to their formation processes.
	The distribution of warm dust emission is investigated towards the 
	central region of the cluster, showing the signature of the impact 
	of massive stars within the cluster region. 
        Radio continuum clumps powered by massive B-type stars 
	(age range $\sim$ 0.07-0.12 Myr) are traced, which are located away 
	from the center of the  stellar cluster NGC 6910 
	 (age $\sim$ 4.5 Myr). 
	Based on the values of different pressure components 
	exerted by massive stars, the photoionized gas associated with the 
	cluster is found to be the dominant feedback mechanism in the cluster.
	Overall, the massive stars in the cluster might have triggered the 
	birth of young massive B-type stars in the cluster. This argument
	is supported with evidence of the  observed age gradient between 
	the cluster and the powering sources of the radio clumps.

	\end{abstract}



	\section{Introduction}

	Massive stars ($\geq$8 M$_\odot$) are regarded as powerful agents which can significantly 
	affect their host molecular cloud through their
	photoionized gas and/or stellar winds. The impact of energetic feedback from massive stars 
	can trigger new generation of young protostars.
	However, understanding the feedback mechanism of massive OB stars is still under 
	debate \citep{2007ARA&A..45..481Z, 2014prpl.conf..149T}.
	 In this context, young open clusters (age $<$ 10 Myr)
        are thought to be a unique laboratory for understanding 
	the processes of star-formation
	as they harbor  both low-mass and high-mass stars of very young ages.
	Young open clusters, just formed from the gravitationally 
	bound molecular clouds and still embedded in the parent nebulous regions, 
	contains dust and gas. 
	The study of the ionized gas, dust (cold and warm) emission, 
	and molecular gas can give us observational clues 
	about the physical processes that govern their formation 
	\citep{2010ARA&A..48..339B, 2015A&A...582A...1D}.

	Furthermore, the investigation of young open clusters offers to study 
	the initial mass 
	function (IMF) of stellar objects, which is an important 
	statistical tool to understand the 
	formation of stars \citep[][and references therein]{2017MNRAS.467.2943S,2017ApJ...836...98J,2018AJ....155...44P}. 
	It is supported with the fact that young open clusters host a broad mass 
	range of cluster members, which can also be used to quantify the 
	relative numbers of stars 
	in different mass bins and to constrain the IMF.
	
	A few examples exist in literature 
	\citep[see e.g.][]{2001A&A...374..504P, 2005MNRAS.358.1290P,2007MNRAS.380.1141S,
	2008AJ....135.1934S, 2017ApJ...836...98J}, showing change 
	in the slope of the mass function (MF) as a function of radial 
	distance from the cluster center
	in a sense that the central region has more number of massive stars as
	compared to the outer regions.
	 Hence, in cases of young clusters, it indicates the imprint of the 
	 star-formation process,
	while in old clusters it may be due to dynamical evolution of the clusters.
	Massive systems sink towards the center, allowing to gain more potential energy which 
	heats the cluster. The time scale for this mass segregation to complete is 
	not very well known. It is considered as an active area of research, especially 
	because of the need to understand trapezium-like sub-systems
	in star clusters \citep{2000ASPC..211...43M}, and the associated implications for the 
	formation mechanisms of massive stars \citep{1998MNRAS.298...93B}.

	Therefore, with the aim to investigate the stellar IMF as well 
	as the physical processes
        governing the interaction and feedback effect of massive stars in their vicinity,
	we have selected a promising young cluster NGC 6910.
	This cluster is believed to host 
	at least ten massive stars of spectral 
	type B2V-O9V \citep[e.g.,][]{2008hsf1.book...36R}.
	 However, the rich population of massive stars in this cluster and their 
	effects on the surrounding field are largely 
	unexplored and  deserve a systematic study. 
	To the best of our knowledge, no comprehensive 
	observational investigation of a large-scale area around NGC 6910 
	is available in the literature. 
	In order to compute the age and distance of the cluster, 
	 in this paper we present new deep, wide-field, and 
	multiband $(UBV(RI)_c)$ photometry 
	around NGC 6910. Furthermore, we have also examined the distribution of massive stars, 
	ionized gas, and warm dust emission in the cluster using multiwavelength
	data sets. 
	
	The structure of the paper is as follow. In Section 2, a brief overview of 
	this region is presented. Section 3 provides details of new optical observations and 
	reduction procedures along with the
	available data sets from various archives. 
	In Section 4, we study the structure of this cluster. In this section, 
	we also discuss the basic parameters of cluster
	(i.e. reddening law, extinction, distance), derived MF slope in the region, 
	and explored the physical environment around the cluster including 
	feedback effects from the massive stars.
	Finally, Section 5 summarizes the various results.

	\section{Overview of the NGC 6910 cluster}
	
	The NGC 6910 cluster was discovered in 1786 by
	William Herschel and many photometric studies of the cluster members have been presented
	later \citep[for details, cf.][]{2008hsf1.book...36R}.
	The distance to NGC 6910 amounts to about 1.5 kpc \citep{1963MNRAS.127...45D,
	1971A&AS....4..241B, 1991MNRAS.249...76B, 1999AstL...25....7D}, 
	placing it behind the Cygnus Rift, within the Local (Orion) spiral arm of the Galaxy. 
	In consequence, the average color-excess of the cluster
	members, $E(B-V)$, is found to be of the order of 1 mag and varies across the cluster
	\citep{1976AJ.....81.1125T}. 
	The age of the cluster was estimated to be in the range between 5 Myr
	and 10 Myr \citep{1963MNRAS.127...45D, 1976ApJS...30..451H,  1991MNRAS.249...76B, 
	2000AJ....119.1848D}.
	
	\begin{figure*}
	\centering
	\includegraphics[width=0.55\textwidth]{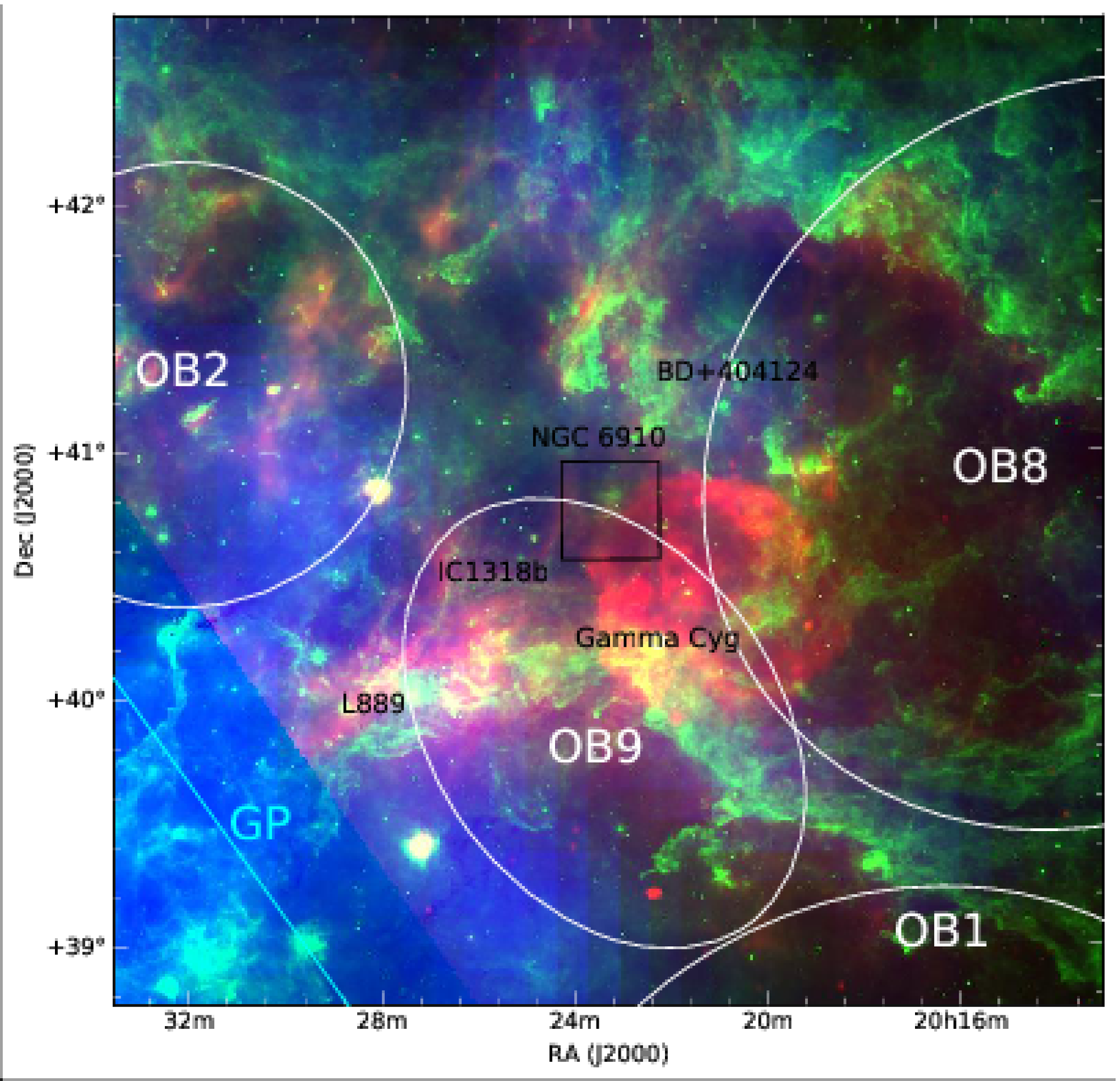}
	\caption{\label{color1} 
	Color-composite image of the $4\times4$ degree$^2$ FOV
	of Cygnus X region containing NGC 6910 (black box) obtained by using 
	1.4 GHz (red, CGPS), 12 $\mu$m (green, $WISE$) 
	and 115 GHz (blue, \citet{2001ApJ...547..792D}) images.
	Locations of different OB associations (white ellipses) are also shown in the figure 
	\citep[cf.][]{1978ApJS...38..309H, 2006A&A...458..855S, 2007A&A...474..873S}.}
	\end{figure*}

	The NGC 6910 cluster is part of a complex of actively star-forming
	molecular clouds and young clusters, the Cygnus X region, which is 
	extended over $\sim$7$^\circ$ $\times$ 7$^\circ$ area and
	located at a distance of about 1.7 kpc \citep{2008hsf1.book...36R}.  
	Several OB stars in Cygnus X have been grouped into 
	nine OB associations by \citet{1978ApJS...38..309H} and the famous Cyg OB2 
	association is the most massive among them, which contains several 
	thousand OB stars and is analogous to the
	young globular clusters in the large Magellanic 
	cloud \citep{1966PROE....5..111R, 1991AJ....101.1408M}.
	OB associations in Cygnus X region are among the largest groups of O-type stars 
	 known in our Galaxy, and can strongly influence their entire surrounding field.
	In Figure \ref{color1}, we show the 
	color-composite of the $4^\circ\times4^\circ$ field-of-view (FOV) of 
	Cygnus X region containing NGC 6910 (see a solid black box) 
	obtained  by using 1.4 GHz Canadian Galactic Plane Survey (CGPS) image, 
	12 $\mu$m Wide-field Infrared Survey Explorer ($WISE$) image, and 115 
	GHz image \citep{2001ApJ...547..792D}.
	The approximate location of OB associations are also shown as white 
	ellipses \citep[cf.][]{1978ApJS...38..309H, 2006A&A...458..855S, 2007A&A...474..873S}.
	The images at 1.4 GHz and 115 GHz represent the distribution of 
	the ionized emission and the molecular gas emission, respectively, while the $WISE$ 12 $\mu$m 
	image covers the prominent polycyclic aromatic hydrocarbon (PAH) features 
	at 11.3 $\mu$m, indicative of photo-dissociation regions (or photon-dominated regions, or PDRs).
	The entire complex seems to contain PAH features produced under the influence of massive stars.
	The cluster NGC 6910 is spatially seen at the border of a dominate circular red 
	region called as `$\gamma$ Cygni 
	supernova remnant (SNR)' \citep[G78.2+2.1; see Figure 1 in][]{2013NuPhS.239...70T}.
	The $\gamma$ Cygni SNR is characterized as a typical shell-type SNR 
	\citep[e.g.,][]{2002ApJ...571..866U,2019ApJ...878...54P}, 
	and is located at a distance of $\sim1.5 $ kpc
	\citep{1980A&AS...39..133L}. 
	A Herbig Ae/Be star BD+40$^\circ$ 4124 \citep{2012A&A...542L..14S} is also seen 
	towards the north-west direction of this cluster.
	The NGC 6910 cluster is located in the outskirts of the IC 1318 b/c bright nebula,
	somewhat within the Cygnus OB9 association of 30\ pc $\times$40\ pc size containing 
	numerous massive young stars. 
	The IC 1318 b/c regions are part of a single, giant H\,{\sc ii}
	region, prominent in the radio domain \citep{1981A&A...101...39B} and 
	bifurcated by a massive, highly structured, dust 
	lane `L889' \citep{1977ApJ...218..133D, 1983A&A...121...69W}.
	A possible ionizing source of the IC 1318 b/c nebula is an  O9V
	type star names as `GSC 03156-00657' \citep{1978AZh....55.1320A,1980A&A....89..239A}.
	However, this star is not a member of NGC 6910, because 
	it is located away
	from the cluster center. 
	The precise relation between the NGC 6910 cluster and the H\,{\sc ii} region is unclear.
	 Also, this cluster has  been known to contain at least
	40 stars showing H$\alpha$ in emission, around 12 pre-main sequence (PMS) stars, and 10 massive stars 
	\citep{1990Afz....32..169M, 1991SvA....35..229S,2000AJ....119.1848D,2007A&A...472..163K,2008hsf1.book...36R},
	which makes it an ideal site to investigate star-formation activities.

	\begin{figure*}
	\centering
	\includegraphics[width=0.55\textwidth]{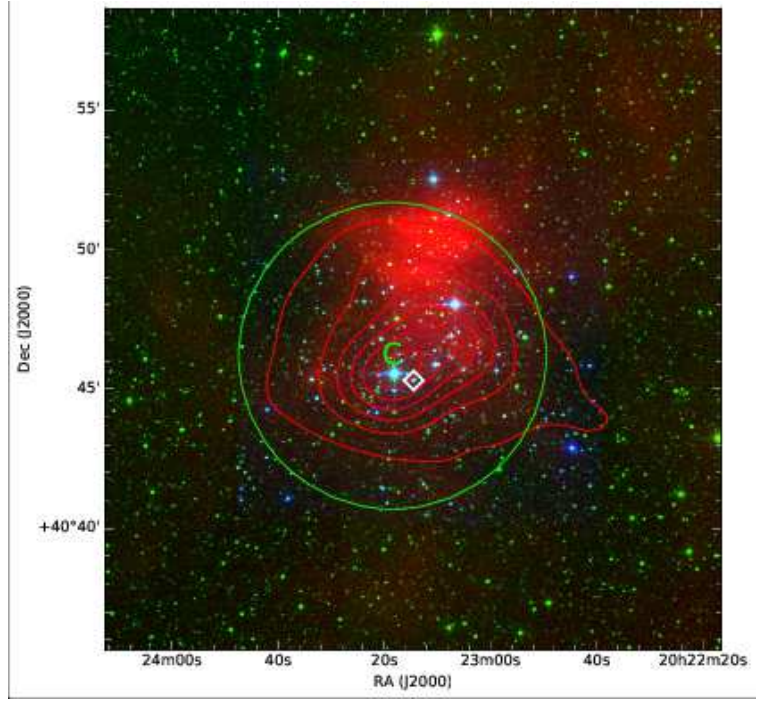}
	\caption{\label{color2} 
	Color-composite image obtained by using the 22 $\mu$m (red, $WISE$), 
	2.2 $\mu$m (green, $2MASS$) and 0.55 $\mu$m (blue, present study)
	images for an area of $\sim22\times 23$ arcmin$^2$ around NGC 6910 cluster. 
	Red contours are the surface density contours, whereas the green circle 
	shows the cluster boundary (cf. Section 4.1.1).
	White diamond is the location of massive O9.5 star (BD+40 4148) reported in this 
	region \citep{2008hsf1.book...36R}.} 
	\end{figure*}

	\section{Observation and data reduction}

	\subsection{Optical data}

	The optical CCD $UBV{(RI)}_c$  photometric data of the NGC 6910 region, 
	centered at $\alpha_{J2000}$: 20$^h$23$^m$12$^s$, $\delta_{J2000}$: +40$^\circ$46$^\prime$42$^\prime$$^\prime$; 
	$l=78^\circ.683$ and $b^\circ=2.013$,
	were acquired 
	by using the $2048\times 2048$ pixel$^2$ CCD camera mounted 
	on the f/13 Cassegrain 
	focus of the 104-cm Sampurnanand telescope of Aryabhatta Research Institute of 
	Observational Sciences (ARIES), Nainital, India. 
	In this set up, each pixel of the CCD corresponds to $0.37$ arcsec and 
	the entire chip covers a FOV of $\sim 13\times13$ arcmin$^2$ on the sky. 
	We have carried out observations of this cluster in four pointings covering 
	a total FOV of $22\times23$ arcmin$^2$
	as shown in Figure~\ref{color2}.
	To improve the signal to noise ratio (SNR), the observations were carried out in 
	the binning mode of $2\times2$ pixels. 
	The read-out noise and gain of the CCD are 5.3 $e^-$ and 10 $e^-$/ADU 
	respectively. The average FWHMs of the star images were $\sim3$ arcsec. 
	A number of bias frames and twilight-flat frames were also taken during 
	observations. A number of short and deep (long) exposure frames 
	were taken to observe both bright and faints stars in the field.  
	The complete log of the observations is given in Table~\ref{log}.

\begin{figure*}
	\centering
	\includegraphics[width=0.55\textwidth]{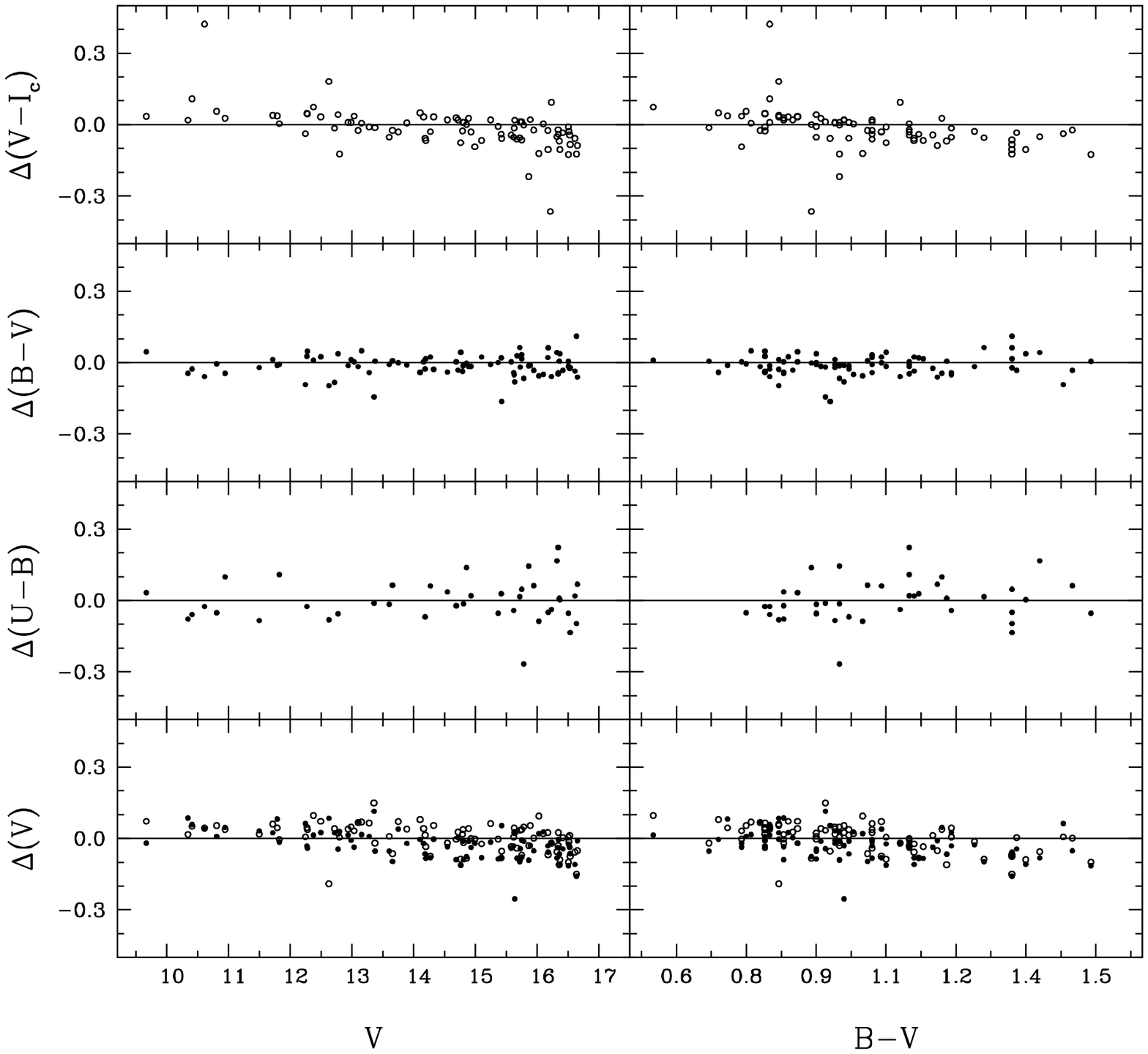}
	\includegraphics[width=0.40\textwidth]{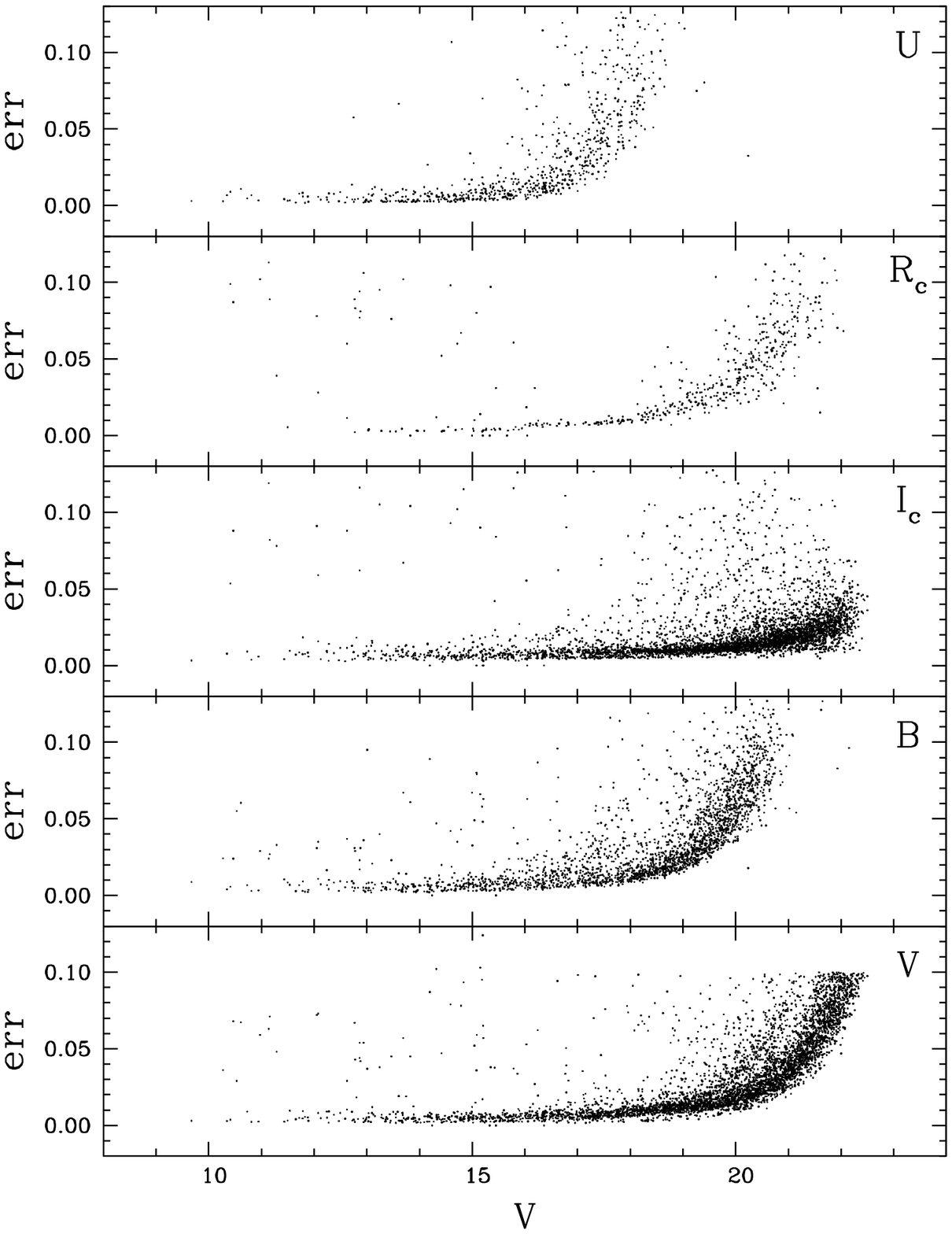}
		\caption{\label{resd} 
	{\it Left panel}: Comparison of the present CCD photometry with those 
	available in literature as a function of present $V$ magnitudes
	and $(B-V)$ colors.
	 Dots and circles represent data from \citet[][$UBV$]{2000AJ....119.1848D} and \citet[][$VI$]{kola1234}, respectively.
	{\it Right panel}: Photometric errors as a function of $V$ magnitude in different filters.
		}
	\end{figure*}

	The CCD data frames were reduced by using the computing 
	facilities available at the Center of Advanced Study, 
	 Department of Physics, Kumaun University, and ARIES, both located in
	Nainital, India.
	Initial processing of the data frames was done by 
	using the IRAF\footnote{IRAF is distributed by National Optical Astronomy
	Observatories, USA} and ESO-MIDAS\footnote{ESO-MIDAS is developed and
	maintained by the European Southern Observatory.} data reduction packages. 
	Photometry of the cleaned frames were carried out by using DAOPHOT-II 
	software \citep{1987PASP...99..191S}.
	The point spread function (PSF) was obtained for each frame by using 
	several isolated stars. 
	Magnitudes obtained from different frames
	were averaged. When brighter stars were saturated on deep exposure frames, 
	their magnitudes were taken from short exposure frames.
	 We used the DAOGROW program \citep{1990PASP..102..932S} 
	for construction of an aperture growth curve 
	required for determining the difference between the 
	aperture and profile-fitting magnitudes.
	 Calibration of the instrumental magnitudes to the standard system was done
	by using the procedures outlined by \citet{1992ASPC...25..297S}.
	The broad-band $UBV{(RI)}_c$ observations of the NGC 6910 region were 
	standardized by observing stars in the 
	SA98 field \citep{1992AJ....104..340L} centered at 
	$\alpha_{J2000}$: 06$^h$52$^m$12$^s$, $\delta_{J2000}$: 
	-00$^\circ$19$^\prime$17$^\prime$$^\prime$.
	The calibration equations derived by using the standard stars in the 
	SA98 field by the least-squares linear regression are as follows:\\

	\begin{equation}
	u= U + (6.771\pm0.003) -(0.009\pm0.002)(U-B) + (0.677\pm0.005)X
	\end{equation}

	\begin{equation}
	b= B + (4.548\pm0.002) +(0.008\pm0.001)(B-V) + (0.407\pm0.003)X
	\end{equation}

	\begin{equation}
	v= V + (4.149\pm0.002) -(0.046\pm0.002)(V-I_c) + (0.256\pm0.003)X
	\end{equation}

	\begin{equation}
	r_c= R_c + (4.046\pm0.002) -(0.013\pm0.001)(V-R_c) + (0.190\pm0.003)X
	\end{equation}

	\begin{equation}
	i_c= I_ + (4.559\pm0.004) -(0.017\pm0.001)(V-I_c) + (0.135\pm0.006)X 
	\end{equation}

	\noindent
	where $U,B,V,R_c$ and $I_c$ are the standard magnitudes 
	and $u,b,v,r_c$ and $i_c$ are the instrumental aperture magnitudes, 
	which are normalized per second of exposure time and $X$ is the airmass.
	 In the cluster field, we have generated secondary standards
	by applying the above equations to the stars which were observed 
	in the same night as the standard field. Then, we calibrated all the
	stars in different subregions of the cluster field which were
	observed during different nights by applying the off-set between the instrumental 
	and standard magnitudes of the secondary standards.
	We have carried out a comparison of the present calibrated data with those CCD 
	data ($V<$17 mag) present in the literature i.e, 
	 \citet[][$UBV$]{2000AJ....119.1848D} and \citet[][$VI$]{kola1234}.
	The difference $\Delta$ (present $-$ literature) as a 
	function of present $V$ magnitudes and $(B-V)$ colors 
	is shown in the left panel of Figure~\ref{resd}.
	 Although, there is some trend
	in the difference of $V$ mags and $V-I_c$ colors 
	in the range of 11 to 16 mag in $V$ band, but the scatter is small
	and the effect will be minimal to the scientific results of this study.
	The comparison indicates that the magnitudes and colors obtained in 
	the present work are in fairly agreement with those available in literature.
	 The typical DAOPHOT errors in different bands  as a function
	of $V$ magnitudes are shown in the right panel of Figure \ref{resd}. It can be seen
	that the errors become large ($>$0.1 mag) for fainter magnitudes
	and were not used in the present analysis. 
	In this study a total of 4638 sources have been identified with detections at 
	least in $V$ and $I_c$ bands and having photometric error less than 0.1 mag 
	upto $V\simeq22$ mag.

	\begin{figure*}
	\centering
	\includegraphics[width=0.48\textwidth]{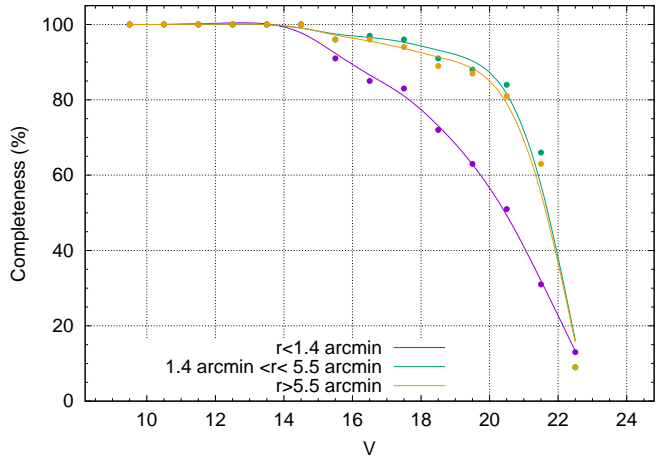}
	\caption{\label{cftf} 
	 Completeness factor as a function of $V$ magnitude
         derived from the artificial star experiment {\it ADDSTAR}.}
	\end{figure*}

	The above optical photometry which will be further used for our analysis, 
	can be incomplete due to various reasons, e.g., 
	nebulosity, crowding of the stars, detection limit etc. In particular it is 
	very important to know the completeness limits in terms of 
	mass to derive correct MF slopes.        
	The IRAF routine ADDSTAR of DAOPHOT II was used to determine the 
	completeness factor (CF). In this method, artificial stars of known magnitudes 
	and positions are randomly added in the original frames and then these 
	artificially generated frames are re-reduced by the same procedure as used 
	in the original reduction. The ratio of the number of stars recovered to 
	those added in each magnitude gives the CF as a function of magnitude.
	We followed the procedure given by \citet{1991A&A...250..324S}. We added
	artificial stars to both $V$ and $I$ images in such a way that they have 
	similar location geometrically but differ in $I$ brightness according to mean
	$(V-I)$ colors of the main sequence stars.
	The luminosity distribution of artificial stars are in such a way that more stars 
	are inserted at fainter magnitude bins.
	A number of independent sets of artificial stars are inserted into a given
	data frames for the determination of the CF. 
	In all about 15\% of the total stars are added so that crowding
	characteristics of the original frame do not change significantly
	\citep[see,][]{1991A&A...250..324S}. 
	The minimum value of the CF in $V$ and $I$ bands is used
	to correct the data incompleteness \citep[see,][]{1991A&A...250..324S}.
	The CF for different regions of the NGC 6910 cluster are shown in 
	Figure \ref{cftf}.
	As expected, we found that the incompleteness of the data increases 
	with increasing magnitude and increasing stellar 
	density (i.e. towards core of cluster, cf. Section 4.1.1). 
	In this study, our data in the cluster region is found to 
	be 80\% complete upto 20.6 mag in
	$V$ band, corresponding to a mass completeness limit of 0.8 M$_\odot$ for 
	 observed distance of 1.72 kpc and $E(B-V)=0.95$ mag (cf. Section 4.1.4). 

	\subsection{Archival data sets}

	We have also used the archival near-infrared (NIR), mid-infared (MIR) and radio 
	data of the selected region as observed in optical bands for our analysis. 
	Brief description of these is given in Table~\ref{surveys}.
	The processed {\it Herschel} temperature and column density ($N(\mathrm H_2)$) 
	maps (resolution $\sim$12$''$) have been utilized in this work, which were 
	downloaded from the publicly available 
	site\footnote[1]{http://www.astro.cardiff.ac.uk/research/ViaLactea/}.
	These maps were generated as a part of the EU-funded ViaLactea 
	project \citep{2010PASP..122..314M}. 
	The Bayesian {\it Point Process MAPping (PPMAP)} procedure \citep{2015MNRAS.454.4282M} 
	was adopted for producing these {\it Herschel} maps \citep[see also][]{2017MNRAS.471.2730M}.

	\begin{figure*}
	\centering
	\includegraphics[width=0.48\textwidth]{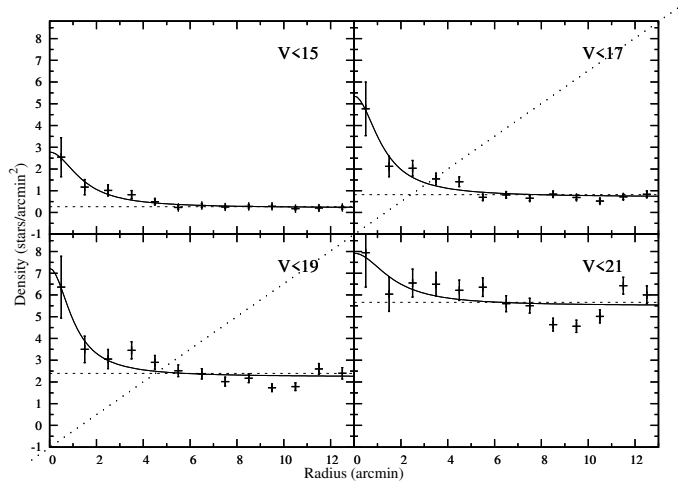}
	\caption{\label{rdp} Radial density profiles of the NGC 6910 cluster at different magnitude 
	levels using the present optical data. The solid curve shows a least-squares 
	fit of \citet{1962AJ.....67..471K}
	profile to the observed data points. The error bars represent 1/$\surd$N  errors.
	The horizontal line indicates the density of field stars.}
	\end{figure*}

	\section{Results and Discussion}

	\subsection{Cluster's physical properties}

	\subsubsection{Extent and structure of the cluster}

	The initial stellar distribution in star clusters may be governed by the
	structure of the parental molecular cloud and also how star-formation 
	proceeds in the cloud. 
	Later evolution of the cluster may be governed by the internal gravitational
	interaction among member stars and external tidal forces due to the
	Galactic disc or giant molecular clouds \citep{2004AJ....128.2306C, 2006AJ....132.1669S}.
	The structure and radius of the NGC 6910 cluster (core and corona regions)  
	can be studied by means of density estimations.
	Since the distribution of stars in a cluster follows a systematic distribution 
	from the cluster to the field region,
	the center of the cluster is estimated by involving a Gaussian kernel with
	the stellar distribution and taking the point of maximum density as the center.
	This was performed for both axes to get the center coordinates of 
	the cluster i.e., $\alpha_{J2000}$: 20$^h$23$^m$18$^s$, 
	$\delta_{J2000}$: +40$^\circ$46$^\prime$12$^\prime$$^\prime$.
	To determine the radial stellar surface density, the cluster 
	was divided into  number of concentric rings.
	The projected radial stellar density in each concentric circle was obtained 
	by dividing the number of stars in each annulus by its area, and this is 
	plotted in Figure~\ref{rdp} for various magnitudes levels.
	The error bars are derived assuming that the number of stars in each 
	annulus follows Poisson statistics.
	The point where the radial density becomes nearly constant and merges with 
	the contaminating field star density  
	(indicated by horizontal dashed lines in the plot) is defined as the radius of the
	cluster `$r_{cl}$' \citep[see also,][]{2006AJ....132.1669S}.
	For almost all magnitude levels, we can determine the $r_{cl}$ of this
	cluster as 5.5 arcmin.
	The observed radial density profile (RDP) of the cluster was parameterized following 
	the approach by \citet{1992AcA....42...29K} in which the projected radial 
	density $\rho(r)$ is described as

	\begin{equation}
	\rho(r) \propto \frac{f_{\circ}}{1+{(\frac{r}{r_c})}^{2}}
	\end{equation}

	where the cluster's core radius `$r_c$' is the radial distance at which the 
	value of projected radial density `$\rho$(r)' becomes half of 
	the central density `$f_0$'. 
	Within uncertainties, the King-model \citep{1962AJ.....67..471K} 
	reproduces well the radial density profile of the cluster  
	at different magnitude levels except for $V<21$ mag. 
	This might be due to the apparent contamination in the cluster region
	from the field stars at fainter magnitudes.
	By fitting the King-model surface density profile to the observed RDP
	of stars with $V\simeq19$ mag having least fitting error, 
	we have found that the core radius of this cluster comes out as 1.4 arcmin.

	To further study the structure of the cluster and stellar density 
	distribution in the region, we generated stellar surface density maps using the 
	nearest neighbor (NN) method as described 
	by \citet{2005ApJ...632..397G}.
	We have taken the radial distance necessary to encompass the 
	$20^{th}$ nearest star detected in optical band ($V<16$ mag) and computed the local 
	surface density in a grid size of $\sim$20 arcsec. 
	Surface density contours in the NGC 6910 region are shown in 
	the Figure~\ref{color2} as red contours,
	whereas, the cluster region as determined by RDP is shown as a green circle. 
	The lowest contour is 1$\sigma$ above the mean of stellar 
	density (i.e. 1.3+2.2 stars/pc$^2$ at 1720\ pc) 
	and the step size is equal to the 1$\sigma$ (2.2 stars/pc$^2$ at 1720\ pc).
	Figure~\ref{color2} reveals that stellar surface 
	density contours correspond to  
	the cluster size (5.5 arcmin) determined by the RDP. 
	The core region of this cluster seems to be elongated.

	\begin{figure*}
	\centering
	\includegraphics[width=8cm]{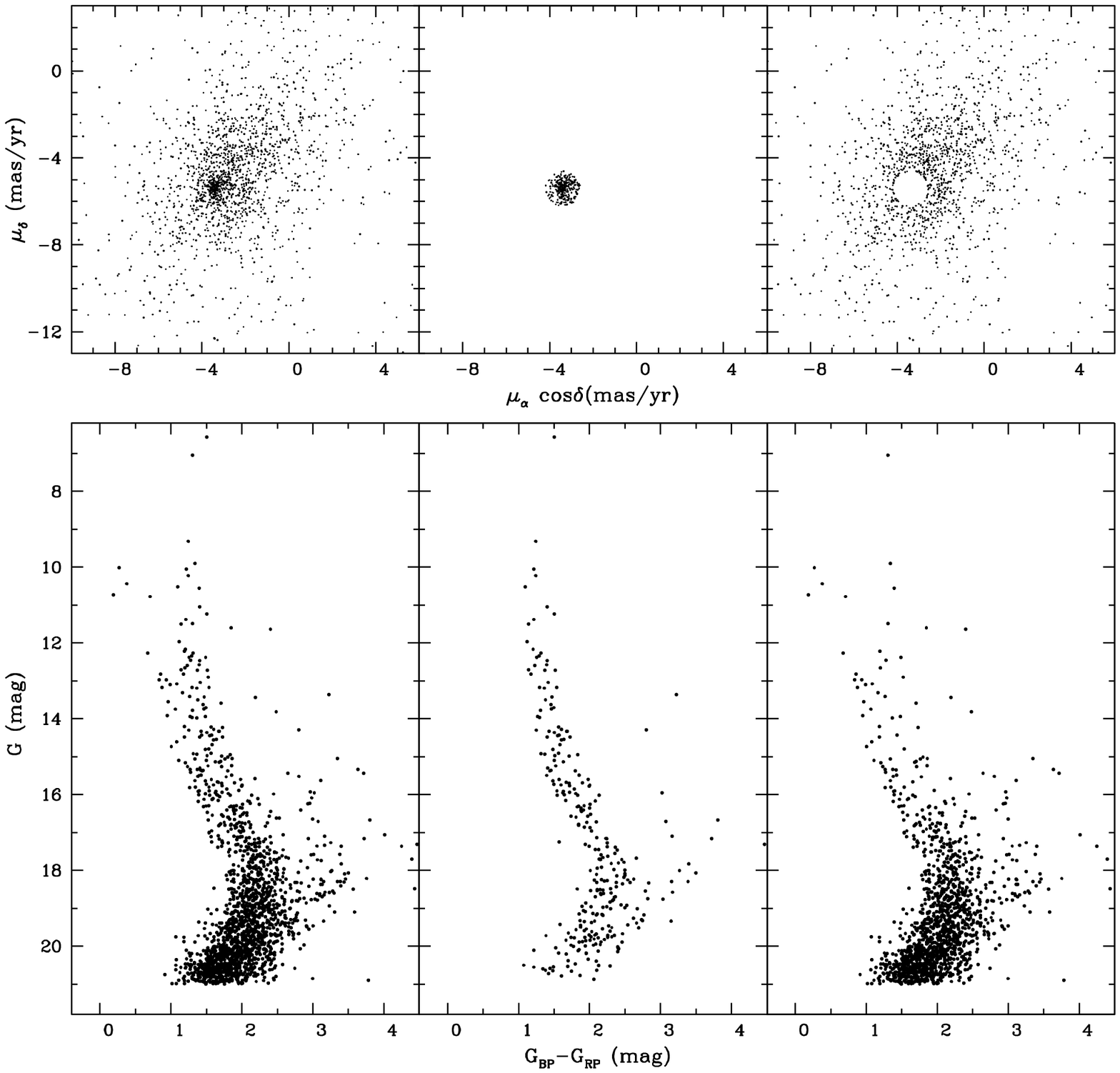}
	\includegraphics[width=7cm,height=8cm]{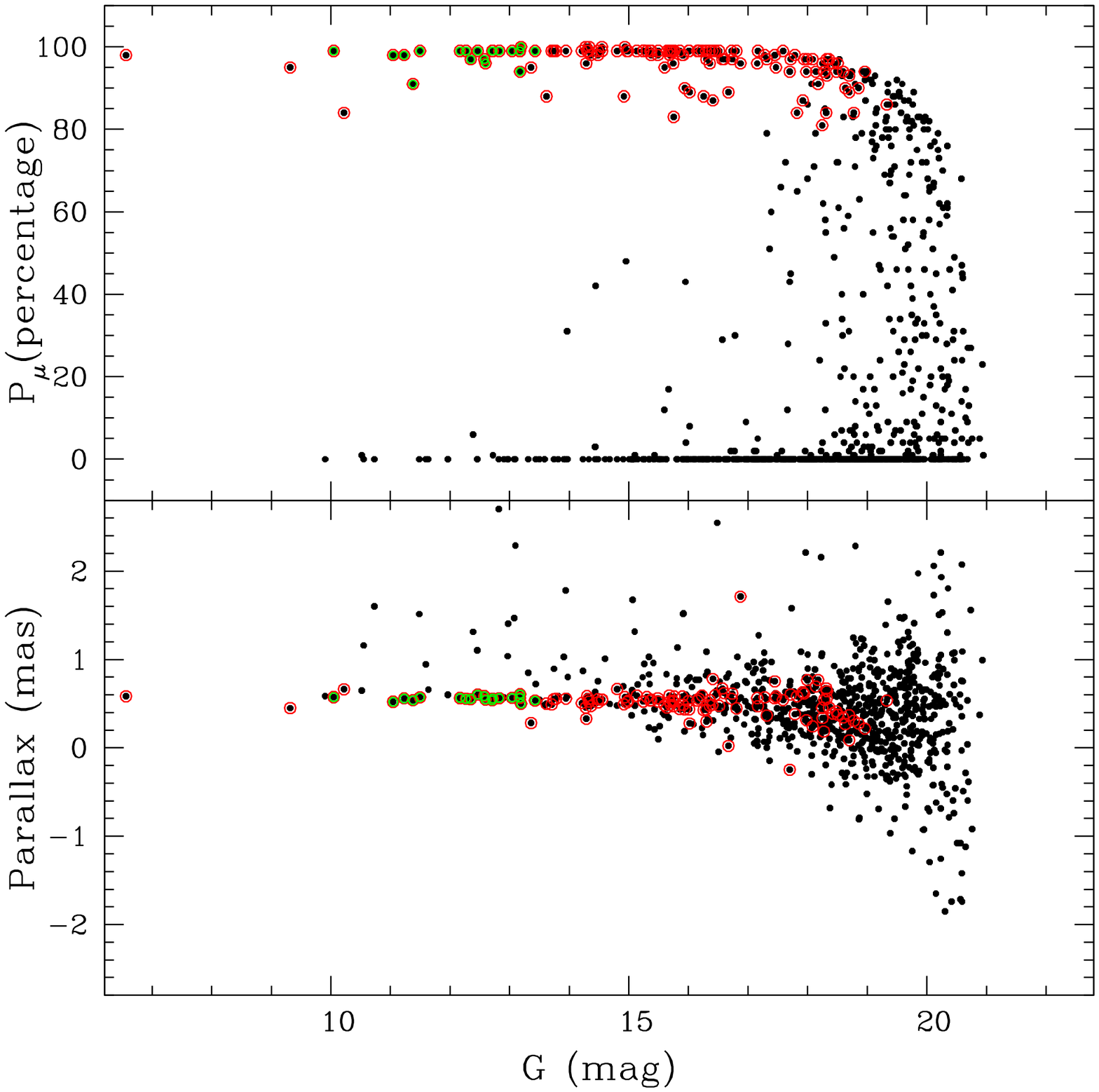}
	\caption{\label{pm1} 
		{\it Left panel:} Proper motion vector-point diagrams (VPDs; top sub-panels) and
	{\it Gaia} DR2 $G$ vs. $(G_{BP} - G_{RP})$ CMDs (bottom panels).
	The left sub-panels show all stars, while the middle and right sub-panels show
	the probable cluster members and field stars.
		{\it Right panel:} Membership probability P$\mu$  and  Parallax values as a 
	function of $G$ magnitude for 
	stars in cluster region.
	Green circles have parallax error less than 0.05 mas and red circles 
	have membership probability
	greater than 80 percent and $G<$20 mag.
	}
	\end{figure*}

	\subsubsection{Cluster membership of stellar sources}

	Membership determination based on proper motion (PM) studies will be useful 
	to carry out astrophysical studies in the region of the cluster. 
	To determine the membership probability, we adopted the method described in 
	\citet{1998A&AS..133..387B} by using $Gaia$ PM data (cf. Table \ref{surveys}).
	This method has been previously used for $\omega$ Centauri
	\citep{2009A&A...493..959B}, NGC 6809 \citep{2012A&A...543A..87S}, NGC 6366 
	\citep{2015A&A...584A..59S} and for NGC 3201 \citep{2017AJ....153..134S}.
	In Figure  \ref{pm1} (left panel), 
	we show the PMs $\mu_\alpha$cos($\delta$) and $\mu_\delta$ vector 
	point diagrams (VPDs: top panels) 
	and the corresponding $G$ versus $(G_{BP} -G_{RP})$ color-magnitude 
	diagrams (CMDs: bottom panels)
	for the stars located within the radius of the cluster i.e., $r_{cl}<$5.5 arcmin.
	The left sub-panel show all stars, while the middle and right sub-panels show 
	the probable cluster members and field stars respectively. 
	 A tight circular clump can be seen visually at $\mu_{xc}$ = -3.4 
	mas yr$^{-1}$, $\mu_{yc}$ = -5.4 mas yr$^{-1}$  having radius of 0.8 mas yr$^{-1}$
	in the top-left sub-panel of Figure  \ref{pm1}. 
	As we know that the cluster stars have more or less similar PMs,
	this group most probably represents the PMs of cluster stars.
	The chosen radius is a compromise between
	losing cluster members with poor PMs and including field
	stars sharing the cluster mean PM.
	The corresponding a well defined CMD of these most probable cluster 
	members can be seen in the lower-middle sub-panel.
	The remaining stars in the VPD are assigned as field stars which 
	is further demonstrated by the
	broad distribution in their CMD (lower-right sub-panel).
	Few cluster members might be visible in
	this CMD because of their wrong estimation of PMs due to large errors in their values.
	Assuming a distance of 1.74 kpc  \citep{2000AJ....119.1848D} and a
	radial velocity dispersion of 1 km s$^{-1}$ for open clusters \citep{1989AJ.....98..227G},
	the expected dispersion ($\sigma_c$) in PMs would be $\sim$0.12 mas yr$^{-1}$.
	For remaining field stars we have calculated: 
	$\mu_{xf}$ = -2.2 mas yr$^{-1}$,  $\sigma_{xf}$ = 3.9 mas yr$^{-1}$ 
	and $\mu_{yf}$ = -4.6 mas yr$^{-1}$, $\sigma_{yf}$ = 4.3 mas yr$^{-1}$,
	as the mean and standard deviation of their PM values in right accession
	and declination axes, respectively. 
	These values are further used to construct the frequency distributions of cluster
	stars ($\phi_c^{\nu}$) and field stars ($\phi_f^{\nu}$) by using the equations
	given in \citet{2013MNRAS.430.3350Y} and then the value of 
	membership probability (ratio of distribution of cluster stars with all the stars) 
	by using the following equation:

	\begin{equation}
	P_\mu(i) = {{n_c\times\phi^\nu_c(i)}\over{n_c\times\phi^\nu_c(i)+n_f\times\phi^\nu_f(i)}}
	\end{equation}
	where $n_c$ (=0.16) and $n_f$(=0.84) are the normalized numbers of stars for the cluster
	and field region ($n_c$+$n_f$ = 1).

	The estimated membership probability of the {\it Gaia} sources located within 
	the radius of the NGC 6910 cluster region is plotted as a function of $G$ 
	magnitude 
	in Figure \ref{pm1} (right panel). As can be seen in this plot, a 
	high membership probability (P$_\mu >$ 80 percent)
	extends down to $G\sim$20 mag. At fainter magnitudes the probability 
	gradually decreases. In Figure \ref{pm1} (right panel), we have also 
	plotted parallax of the same stars as a function of $G$ magnitude.
	Except few outliers, most of the  stars with high membership 
	probability (P$_\mu >$ 80 percent and $G<$20 mag) are
	following a tight distribution.
	Finally, from the above analysis, we calculate membership 
	probability of 916 stars in
	the NGC 6910 cluster region, and 128 stars were assigned as cluster 
	members based on their high membership probability 
	(P$_\mu$ $>$80 percent and  $G<$20 mag). 
	The details of these cluster members are given in Table \ref{PMT}.

	\begin{figure*}
	\centering
	\includegraphics[width=0.55\textwidth]{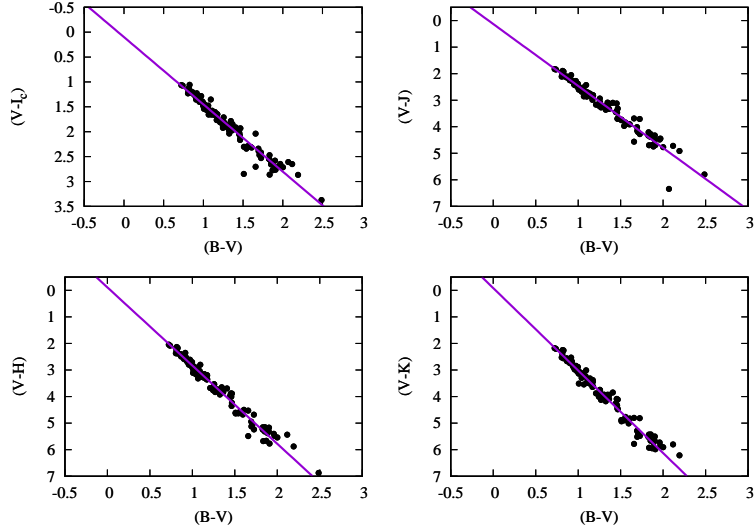}
	\caption{\label{2color} $(V-I_c),(V - J), (V -H), (V - K)$ versus $(B - V)$ TCDs
	for the stellar sources associated with the NGC 6910 region 
	(black dots, P$_\mu >$80\% and $G<20$ mag) and  
	straight lines show the least-square fit to the distribution of stars.}
	\end{figure*}

	\subsubsection{Reddening law in the region}

	The nature of diffuse interstellar medium (ISM) is often 
	characterized by the 
	ratio of total-to-selective extinction, represented by $R_V$ = $A_V$/$E(B-V)$.
	The normal reddening law  for the solar vicinity gives the
	value $R_V$ = 3.1$\pm$ 0.2 
	\citep{2003dge..conf.....W, 1989AJ.....98..611G, 2011JKAS...44...39L},
	but in the case of several star-forming 
	regions having an unusual distribution of dust sizes, it is found to be
	abruptly high \citep[see e.g.,][]{ 2000PASJ...52..847P, 2008MNRAS.383.1241P, 2012AJ....143...41H, 2013ApJ...764..172P, 2014A&A...567A.109K}.
	To separate the influence of normal extinction produced by
	the general ISM from that of abnormal extinction arising within regions, we 
	used $(V - \lambda)$ versus $(B - V)$ two-color diagrams (TCDs) 
	\citep[cf.,][]{1990A&A...227L...5C, 2000PASJ...52..847P,2003AA...397..191P}, 
	where $\lambda$  indicates one of the wavelengths of the broad-band 
	filters ($J, H, K, I_C$).
	 Figure~\ref{2color} shows the $(V - \lambda)$ versus $(B - V)$ TCDs of 
	all the cluster member stars having optical and NIR observations.
	 The slopes of the least square fit to the distribution of 
	stars in the $(V-I_c),(V-J),(V-H)$ and $(V-K)$ 
	versus $(B-V)$ TCDs are found to be $1.35\pm0.10, 2.33\pm0.12, 2.94\pm0.11$ and 
	$3.10\pm0.09 $, respectively, 
	which are higher by a factor `m'$\sim1.21\pm0.01$ than those found for the general ISM 
	\citep[1.10, 1.96, 2.42 and 2.60; cf.,][]{2003AA...397..191P}.
	This concede a higher value for $R_V$
	\citep[$\sim3.75\pm0.02$, please refer][for detailed description on reddening law estimation]
	{2003AA...397..191P},
	indicating larger grain sizes of the material in this region as compared to the general ISM.
	We have also calculated the $R_V$ value for the stars which 
	are not cluster members  
	(i.e. P$\mu <$ 20 per cent and $V<20$ mag) and found a similar 
	value of $R_V$ ($\sim3.75\pm0.02$) as for cluster member stars.
	This means that, in general, the $R_V$ value is higher in 
	this region.
	Inside dense dark clouds, the coagulation due to grain collision 
	and accretion of ice mantles on grains can change the
	 size distribution leading to higher $R_V$ 
	values \citep{1989ApJ...345..245C}.
	In many star-forming regions, $R_V$ values tend to diverge from the normal value 
	towards the higher ones,
	for example: $R_V$ = 3.7 \citep[][the Carina region]{2014A&A...567A.109K}, 
	$R_V$= 3.3 \citep[][NGC 1931]{2013ApJ...764..172P},
	$R_V$ = 3.5 \citep[][NGC  281]{2012PASJ...64..107S} and $R_V$ = 3.7 
	\citep[][Be 59]{2008MNRAS.383.1241P}.

	\begin{figure*}
	\centering
	\includegraphics[width=0.45\textwidth]{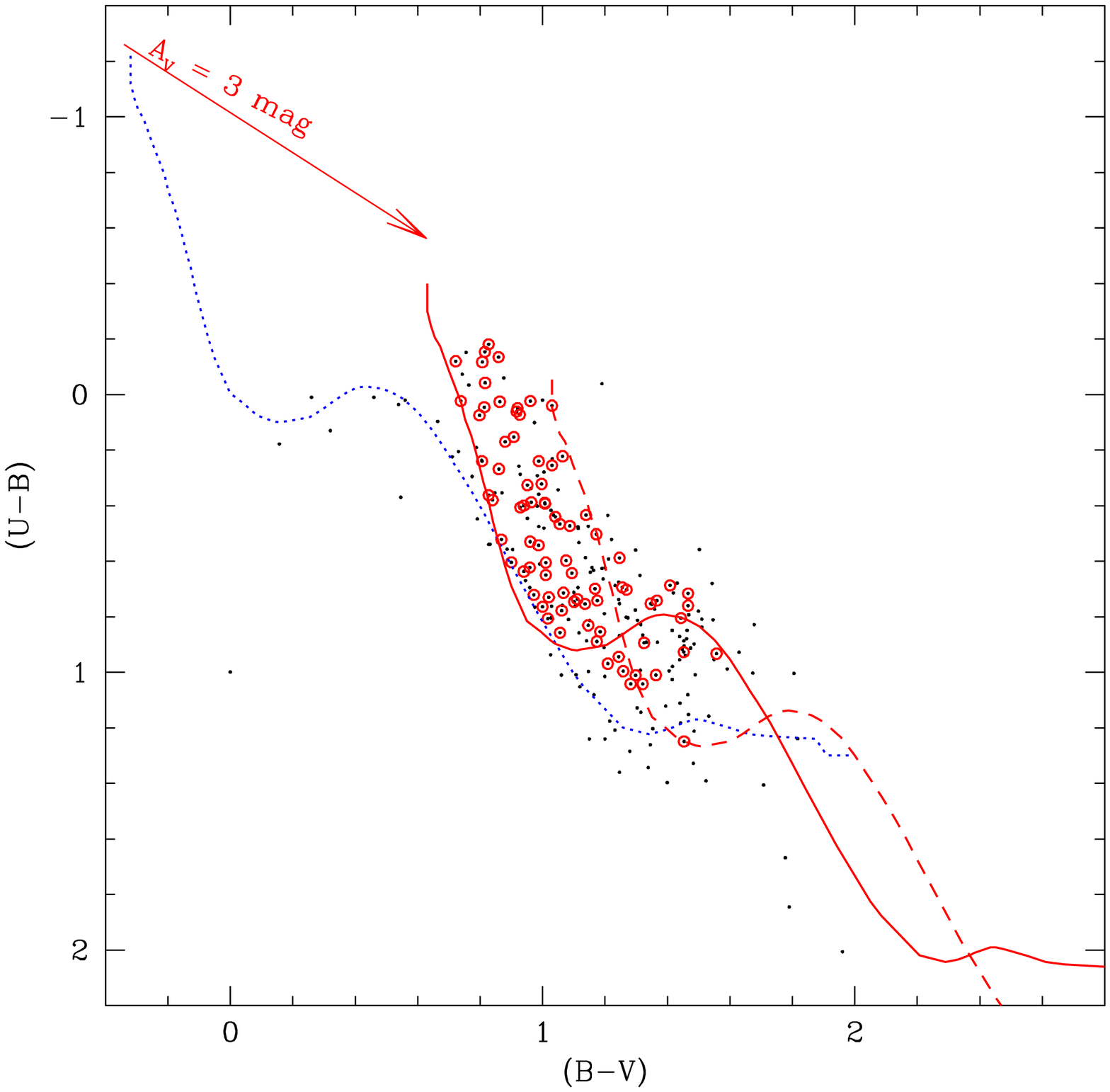}
	\includegraphics[width=0.45\textwidth]{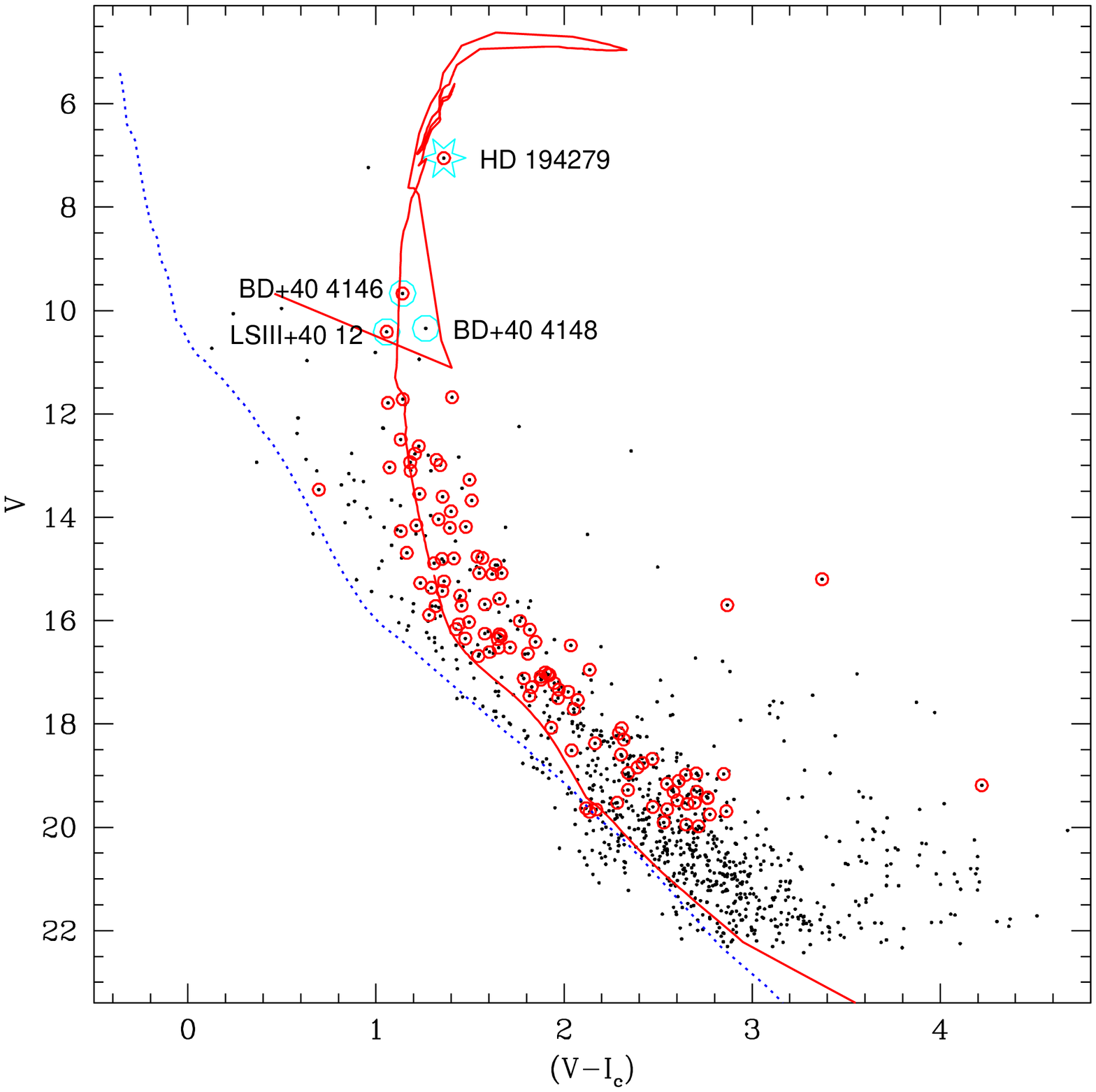}
	\caption{\label{cmd} 
	{\it Left panel}: $(U-B)$ vs. $(B-V)$ TCD for all the optically detected 
	sources in the NGC 6910 region ($r_{cl}<5.5^\prime$).
	Red open circles are cluster member stars identified by their PMs data.
	The dotted blue curve represents the intrinsic Zero Age Main 
	Sequence (ZAMS) for $Z=0.02$ by \citet{2013ApJS..208....9P}.
	The continuous and dotted red curves represents ZAMS shifted along 
	the reddening vector 
	for $E(B-V)$ = 0.95 mag and 1.35 mag, respectively. 
	{\it Right panel}: $V$ vs. $(V-I_c)$ CMD for similar sources.
	 The ZAMS \citep[][blue dotted curve corrected for the distance of 0.8 kpc]{2013ApJS..208....9P}  and 
	post main-sequence isochrone for 4.5 Myr 
	\citep[][solid red curve corrected for the distance of 1.72 kpc and reddening $E(B-V)=0.95$ mag]{2019MNRAS.485.5666P} are also shown. 
	}
	\end{figure*}

	\subsubsection{Extinction, distance and age of the cluster}

	The extinction towards the cluster NGC 6910 can be estimated by using the 
	$(U - B)$ versus $(B - V)$ TCD as shown in 
	the Figure~\ref{cmd} (left panel). In this figure,
	stars inside the cluster region ($r_{cl}$ $<$ 5$'$.5, black dots) along with intrinsic 
	zero-age-main-sequence (ZAMS, blue dotted curve) taken 
	from \citet{2013ApJS..208....9P} are shown.
	We have also over-plotted the probable cluster members stars 
	identified by using the PMs 
	data (cf. Section 4.1.2) as red circles.
	The distribution of the stars shows a large spread along the reddening vector, 
	indicating heavy differential reddening in this region.
	It reveals two different populations, one (mostly black dots) distributed along the ZAMS
	and another (mostly red circles) showing a large spread in their reddening value.
	The former having negligible reddening must be the foreground population and the 
	latter could be member stars.
	If we look at the MIR images of this region (Figure~\ref{color1}),
	we see several dust lanes along with enhancements
	of nebular emission at many places. Both of them are likely responsible for the large 
	spread of reddening in the NGC 6910 region.
	The ZAMS is  shifted along the reddening vector with a slope of
	$E(U - B)/E(B - V)$ =  {\it $0.72\times1.21$ }(corresponding to $R_V$ $\sim$ 3.75, cf. 
	Section 4.1.3)
	to match the distribution of stars showing  minimum reddening among the 
	member population. 
	 Only those stars were chosen for reddening analysis if their position in the 
	TCD indicated that their spectral type is A or earlier.
	This choice  was dictated by several factors, such as, 
	metallicity, distribution of binary stars, rotation,  PMS stars
	and error in photometry 
	\citep[see for more detail, cf.][]{1974ASSL...41.....G,1994ApJS...90...31P}.
	The cluster foreground reddening value, $E(B-V)_{min}$ thus comes 
	out to be $\sim$0.95 mag
	and the ZAMS reddened by this amount is shown by a red continuous curve.
	The other stars may be embedded in the nebulosity of this region and
	the maximum reddening value, $E(B-V)_{max}$  for them comes out to be 1.35 
	mag (dashed red curve).
	The approximate error in the reddening measurement `$E(B-V)$' is $\sim$ 0.1 mag,
	 and has been determined by the procedure  outlined in \citet{1994ApJS...90...31P}.

	 Photometric distance ($\sim$ 1.5 kpc) of this cluster has been estimated 
	previously in the literature
	\citep{1991SvA....35..229S,1992BaltA...1...31V}.
	Recent distance estimates put this cluster at a distance of 1.6 
	kpc \citep{kola1234} to 1.74 kpc \citep{2000AJ....119.1848D}.
	\citet{2000AJ....119.1848D} have quoted the
        distance and age of this cluster as 1740\ pc and  6.8 Myr, respectively.
	However, as the extinction in the NGC 6910 region is high and apparently anomalous, 
	these distance measurements will be sensitive to the adopted $R_V$ values. 
	We calculated the distance of the member stars of this cluster by using their
	parallax values with good accuracy (error $<$ 0.05 mas)
	from \citet{2018AJ....156...58B}. The mean distance value 
	comes out to be  $1.72\pm0.08$ kpc.
	Distance and age of a cluster can also be derived quite accurately by using 
	the CMD of their main-sequence (MS) 
	member stars \citep[cf.][]{1994ApJS...90...31P,2006AJ....132.1669S,2017MNRAS.467.2943S,2014A&A...563A.117F,2015A&A...576A...6P,2019A&A...623A.108B,2020MNRAS.492.2446P,2020ApJ...891...81P}.
	The $V$ versus $(V-I)$ CMD for stars lying within the cluster 
	region is shown in the right panel 
	of Figure~\ref{cmd}. The probable cluster member stars (cf. Section 4.1.2) are 
	also plotted in the figure by red circles.  
	Here also, the CMD reveals two different populations, one (mostly black dots) for
	foreground stars having almost zero reddening value (near dotted curve) and 
	another (mostly red circles) for the cluster members at higher reddening value 
	and larger distance. The CMD of cluster members displays a few MS 
	stars upto $V$=16 mag 
	and PMS stars at fainter end. 
	The blue dotted curve in the right panel
        of Figure~\ref{cmd} denotes a ZAMS from \citet{2013ApJS..208....9P}, randomly 
	corrected for a distance of 0.8 kpc, 
	matches well with the distribution of foreground stars.
	We have further visually fitted the post-MS isochrone for an age of 4.5 Myr
	from \citet{2019MNRAS.485.5666P} to the
	lower envelop of the distribution of member stars where the bend occurs in the MS.
	This choice of visual fitting was imposed by several factors, such as, 
	distribution of binary stars, rotation and evolutionary effects 
	\citep[see for detail, cf.][]{1974ASSL...41.....G,1994ApJS...90...31P}.
	It matches quite nicely which is corrected for a 
	extinction value `$E(B-V)_{min}$=0.95 mag'
	as derived earlier in this section and a  distance of 1.72 kpc  (solid red curve).
	The locations of massive stars such as HD 194279 
	\citep[B2Ia C, Blue super-giant;][]{2007BaltA..16..311A}, 
	BD+40 4148 \citep[O9.5V;][]{1965ApJS...12..215H}, 
	BD+40 4146 \citep[B1 D;][]{1968PASP...80..290W}
	and LS III +40 12 \citep[B0.5V;][]{2012A&A...543A.101C} 
	are also matching well with the isochrone.
	Therefore, from both parallax and CMD analyses, 
	we have derived the distance and post-MS age of this cluster as 1.72 kpc
	and 4.5 Myr, respectively.
        The approximate error in the age estimation is $\sim$2.5 Myr,
        as has been determined by the procedure  outlined in \citet{1994ApJS...90...31P}.
	 A summary of physical parameters of the cluster is given in Table \ref{ST}.

	\subsubsection{Mass function and dynamical age of the cluster}

	The distribution of stellar masses that form in one star-formation 
	event in a given volume of space is called IMF and 
	together with star-formation 
	rate, it is one of the important statistical tools to study star-formation.
	The MF is often expressed by a power law,
	$N (\log m) \propto m^{\Gamma}$ and  the slope of the MF is given as:
	\begin{equation}
	 \Gamma = d \log N (\log m)/d \log m 
	\end{equation}
	\noindent
	where $N (\log m)$ is the number of stars per unit logarithmic mass interval.
	We have used our deep optical data to generate the MF of different 
	regions of the NGC 6910 cluster 
	as it reaches to the fainter end as compare to the $Gaia$ DR2 data. 
	For this, we have utilized the optical CMDs of the
	sources in the target region and that of the nearby field region of equal area
	and decontaminated the former sources from foreground/background
	stars and corrected for data in-completeness using a statistical subtraction method
	already described in details in our previous papers
	\citep[cf.][]{2007MNRAS.380.1141S,2012PASJ...64..107S,2017MNRAS.467.2943S,
	2008MNRAS.383.1241P,2013ApJ...764..172P,2011MNRAS.415.1202C,2013MNRAS.432.3445J}.

	\begin{figure*}
	\centering
	\includegraphics[width=0.60\textwidth]{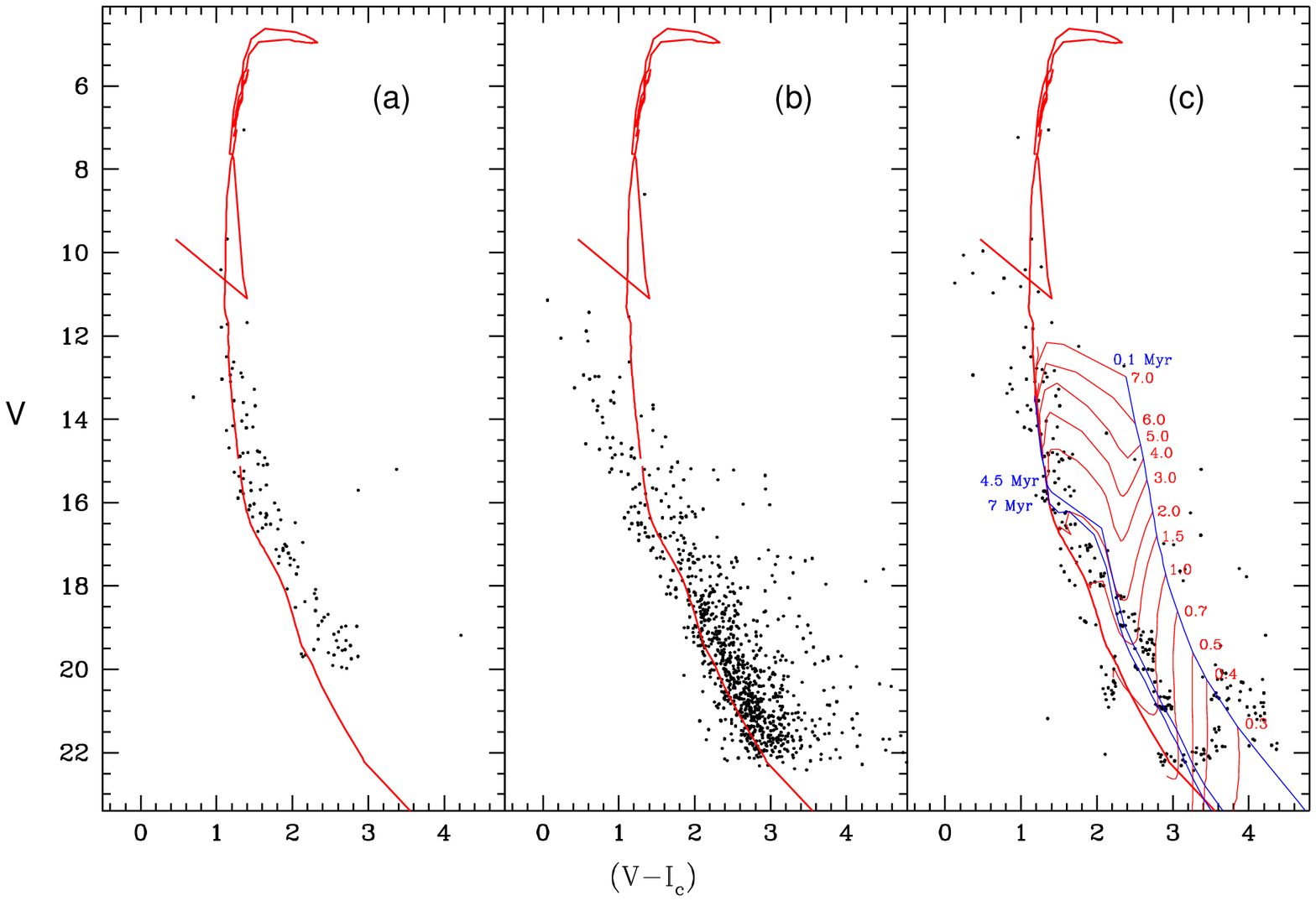}
	\includegraphics[width=5.5cm,height=7.5cm]{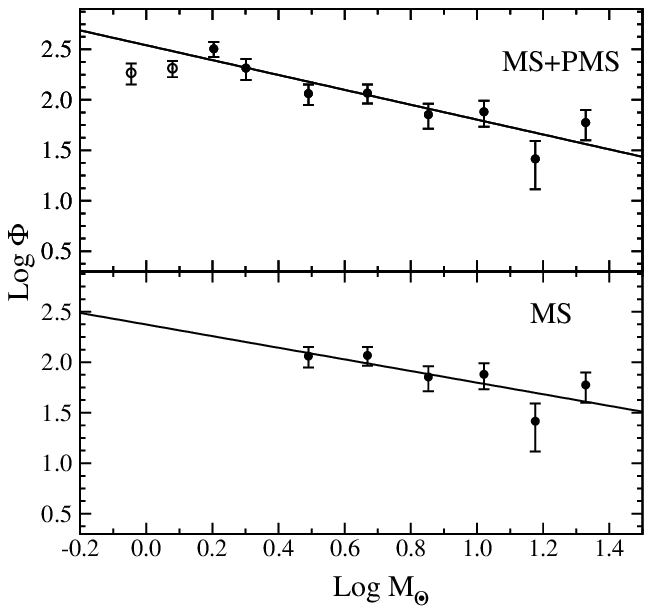}
	\includegraphics[width=0.43\textwidth,angle=0]{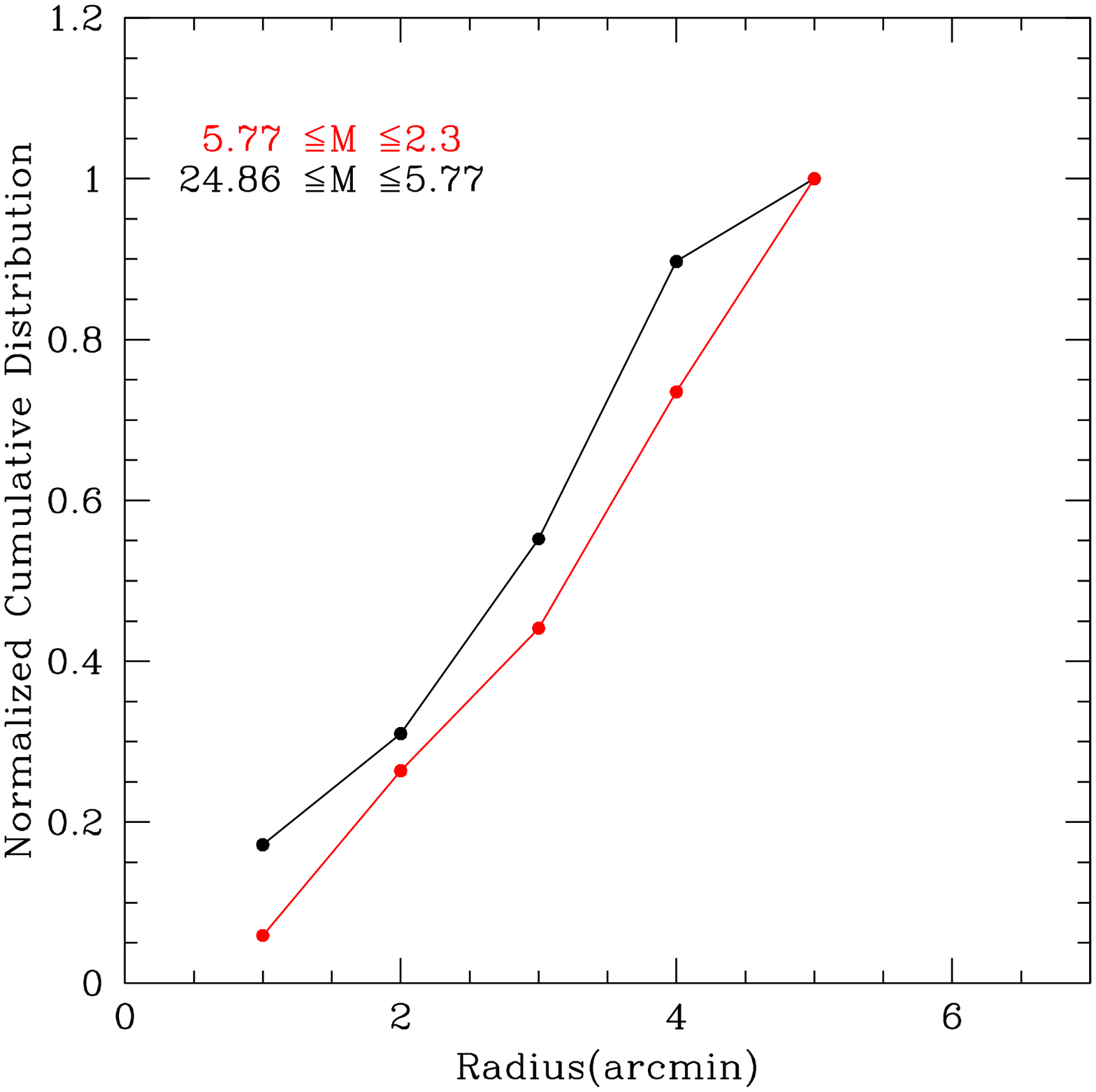}
	\includegraphics[width=0.43\textwidth,angle=0]{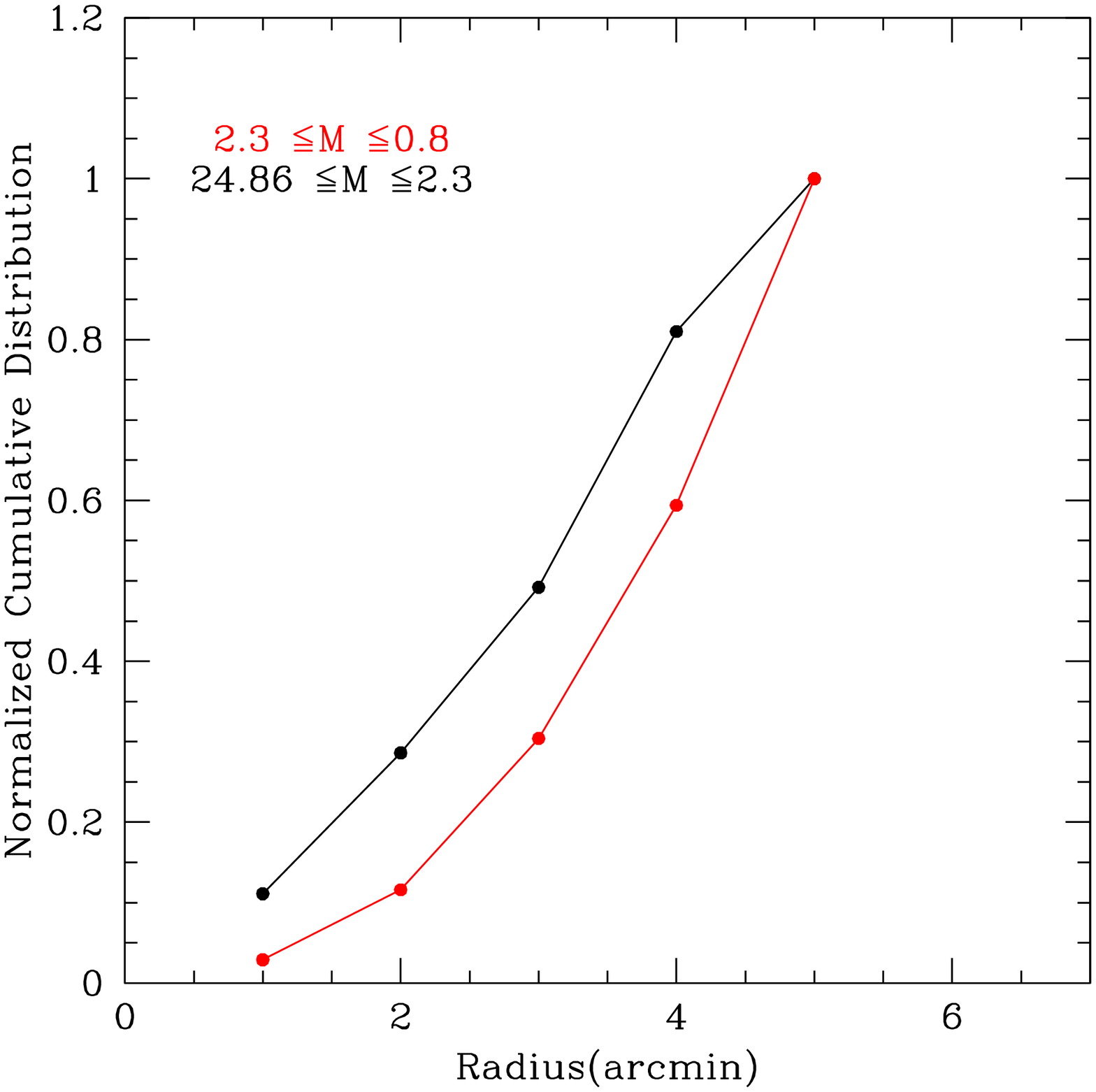}
	\caption{\label{band} {\it Top left}: $V$ vs. $(V-I_c)$ CMD for (a) Stars in the 
	NGC 6910 region,
	(b) Stars in the reference region.
	and  (c) Statistically cleaned CMD overplotted with the isochrone for 4.5 Myr
	from \citet{2019MNRAS.485.5666P} (thick red curve) along with the 
	pre-main sequence (PMS) 
	isochrones of 0.1, 4.5 and 7 Myr (blue curves) and the 
	evolutionary tracks of different masses (red curves) 
	from \citet{2000AA...358..593S} are shown.
	All the isochrones and evolutionary tracks are corrected for the distance 
	of 1.72 kpc and reddening $E(B-V)$ = 0.95 mag.
	{\it Top right}: A plot of MFs for the stars of 
	the NGC 6910. log$\phi$ represents log(d$N$/dlog$m$). The error bars 
	represent $\pm\sqrt N$ error. Continuous curves show a least-squares fit for the given mass 
	range. The upper sub-panel includes stars in the main-sequence (MS) 
	as well as PMS phase
	whereas the bottom sub-panel includes only MS stars.
	{\it Bottom}: Cumulative radial distribution of MS stars (bottom left) and  MS+PMS 
	stars (bottom right) for different mass intervals.  }
	\end{figure*}
	
	As an example, in Figure~\ref{band} (top-left panel) we have shown $V$ 
	versus $(V-I_c)$ CMDs for the stars lying within the 
	cluster region in sub-panel `a' 
	and for those in the reference field region
	(taken as an annular area outside the cluster region 
	having radius $5^\prime.5<r_ {field}<7^\prime.65$) 
	in sub-panel `b'.
	In  sub-panel `c', we have plotted the statistically cleaned $V$ versus $(V-I_c)$
	CMD for the cluster region which is showing the presence of PMS stars.
	Since, at the age of 4.5 Myr, the stars having $V\leq 16$ mag are considered to
	be still on the MS, for these stars the luminosity function was converted into MF
	using theoretical models by \citet{2019MNRAS.485.5666P}. 
	The MF for the PMS stars was obtained by counting the 
	number of stars in various mass bins (shown as evolutionary tracks) having 
	age $\leq7$ Myr (age of cluster i.e. 4.5 Myr + error in age) in Figure~\ref{band} 
	sub-panel `c'. The resulting MF of the cluster region by using
	MS ($2.3<M/M_\odot<24.86$) and MS+PMS ($0.8<M/M_\odot<24.86$) stars are plotted 
	in Figure~\ref{band} (top-right) in upper and 
	lower sub-panels, respectively. Similarly, we have also derived MF 
	slopes `$\Gamma$' for core and corona regions of the cluster using
	both MS and MS+PMS stars and their values are given in Table~\ref{mfslope}.

	 As the NGC 6910 cluster region contains several massive stars,
	we shall study the environment effects due to the presence of high mass stars
	on lower-mass end of the present day MF.
	In Figure~\ref{band} (top-right panel), we can observe that the MF for the cluster region 
	 shows a turn-off at 1.58 M$_\odot$ and the distribution upto this mass limit 
	can be represented by a single power law.
	The MF slopes of different regions of the cluster NGC 6910 are in general 
	shallower than the Salpeter value `-1.35', which indicates the 
	abundance of massive stars in this cluster.
	It has been shown \citep[see e.g.][]{1986FCPh...11....1S, 1998ASPC..142..201S, 
	2002Sci...295...82K, 2003PASP..115..763C, 2005ASSL..327.....C} that, 
	for masses above $\sim$1 M$_\odot$, the MF
	can generally be approximated by a declining power law with a slope similar to
	that found by \citet{1955ApJ...121..161S}.
	To investigate further, we look for the signature of mass-segregation in this cluster
	by checking the change of MF slope from the core region to the outer corona region of 
	this cluster, which is in fact getting steeper in the outer corona region. 
	To evaluate the degree-of-mass-segregation in the cluster, we
	subdivided the samples of cluster stars  into two mass groups 
	and plotted their cumulative distribution with respect to radial distance 
	from the cluster center as shown in Figure~\ref{band} (bottom panels) for both 
	MS (bottom-left panel) and MS+PMS (bottom-right panel) stars.
	Figure~\ref{band} (bottom panels) also reveals the effect of mass-segregation 
	in the sense that relatively massive
	stars tend to lie near the cluster center. The Kolmogrov-Smirnov test
	confirms the above-mentioned mass-segregation at a 
	confidence level better than $\sim$99\%.

	Dynamical relaxation is one of the possible reasons 
	of the segregation of massive stars in the central region of this cluster.
	At the time of formation, the cluster may have a
	uniform spatial stellar mass distribution, however, the spatial stellar mass 
	distribution would change with time as the cluster evolves dynamically. 
	Low-mass stars in a 
	cluster may possess high random velocities because of the dynamical 
	relaxation; consequently, 
	they will try to occupy a larger volume than the high mass stars and move away from the 
	cluster center \citep{1986ApJ...310..613M, 1985IAUS..113..427M}.
	To check whether mass-segregation is primordial or due to dynamical relaxation, 
	we have estimated the dynamical relaxation time, `$T_E$', the time in which 
	individual stars exchange sufficient energy so that their velocity distribution 
	approaches that of a Maxwellian equilibrium. The dynamical relaxation time is 
	given by

	\begin{equation}
	T_E = \frac{8.9\times10^{5}N^{1/2}R_h^{3/2}}{{m}^{1/2}log(0.4N)}
	\end{equation}

	where, $N=215$ is the number of cluster members, $R_h=1.6$ pc is the radius containing 
	half of the cluster mass and $m=4.31 M_\odot$ is the average mass of the cluster stars 
	\citep{1971ApJ...164..399S}. 
	The total number of MS stars and the total mass of MS stars (775 M$_\odot$) in the given 
	mass range are obtained with the help of the MF. This total mass
	of MS stars should be considered as a lower limit to the total mass of the 
	cluster.
	The half mass radius $R_h$ as half of the cluster radius appears a tenable 
	approximation. We used half of the cluster radius ($r_{cl}$) 
	obtained 
	from the optical data as the half-mass radius.
 	The dynamical age of this cluster comes out to be 6.5 Myr
	which is more than the age of this cluster (i.e. 4.5 Myr), suggesting that 
	the dynamics is not fully responsible for the observed mass-segregation and 
	it may be the imprint of star-formation processes itself.

	\subsection{Physical environment around the cluster}

	\subsubsection{Multi-wavelength picture of the region}
	
	Recently available high resolution radio/infrared/sub-millimeter
        observations have helped us to probe deeply embedded star-forming regions,
        and have provided a wealth of new information to probe young stars, gas 
	and dust distribution, ionized gas distribution
        etc, which are very good indicators/tools for star-formation 
	studies \citep[e.g.,][]{2010A&A...523A...6D,2010ApJ...716.1478W,2016ApJ...819...66D,2017ApJ...834...22D}.
	
	\begin{figure*}
	\epsscale{1}
	\plotone{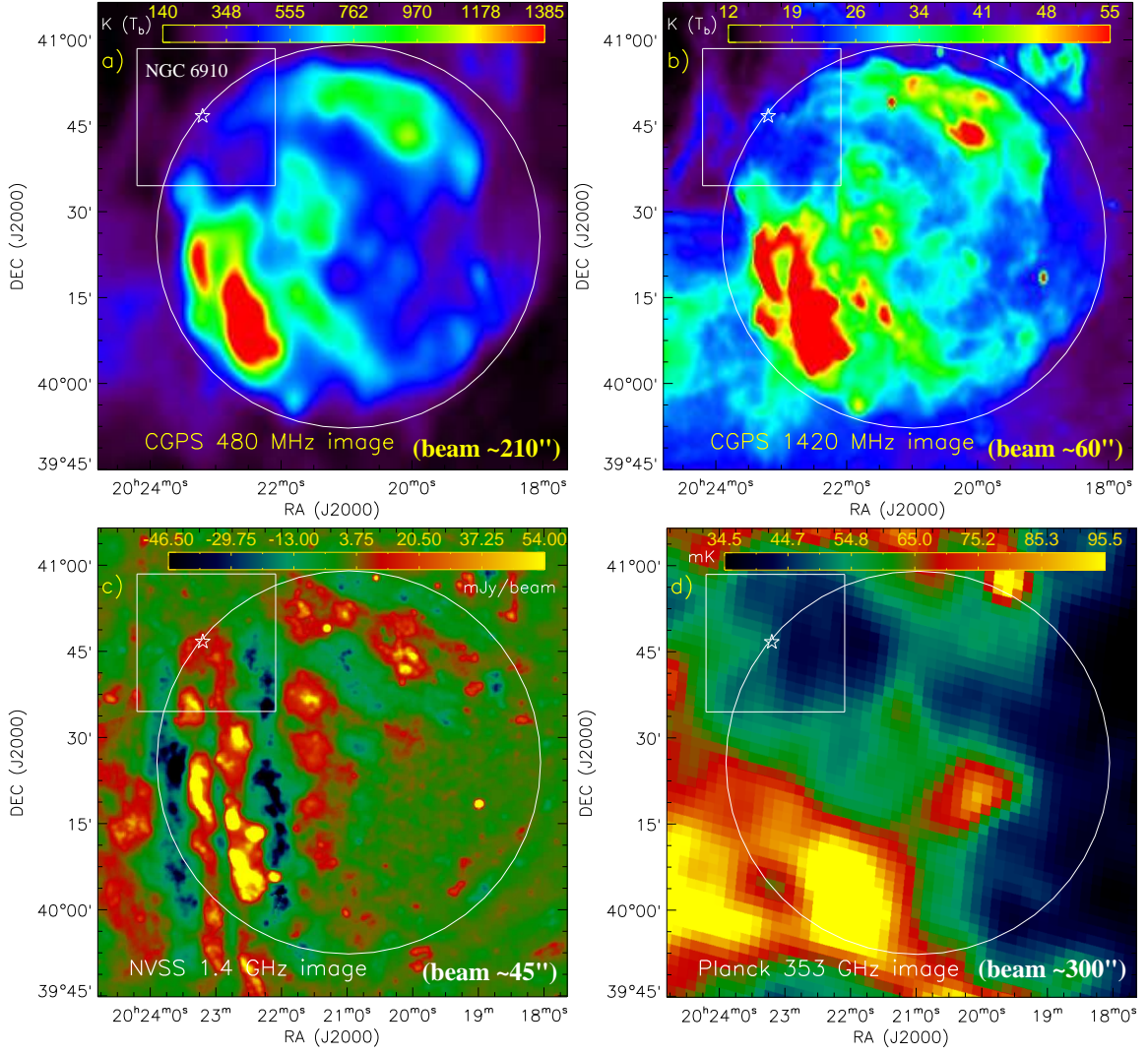}
	\caption{a) CGPS 480 MHz radio continuum image of an area (size $\sim$1$\degr$.36 $\times$ 1$\degr$.36; 
	centered at $\alpha_{2000}$ = 20$^{h}$ 21$^{m}$ 15.8$^{s}$, $\delta_{2000}$ = +40$\degr$ 25$'$ 49$''$) containing the site NGC 6910.  
	b) CGPS 1420 MHz radio continuum image.
	c) NVSS 1.4 GHz radio continuum image.
	d) {\it Planck} 353 GHz continuum image.
	In each panel, the white box encompasses the area selected in this paper and 
	is shown as a zoomed-in view in Figures~\ref{fig2}a and~\ref{fig2}b,
	a circle indicates an extended spherical-like structure traced in 
	the CGPS 1420 MHz continuum image (see Figure~\ref{fig1}b) and 
	a star symbol represent the central position of the stellar cluster.}
	\label{fig1}
	\end{figure*}
	
	In Figures~\ref{fig1}a and~\ref{fig1}b, the large-scale environment of the cluster NGC 6910 
	is shown using the CGPS 480 and 1420 MHz continuum images, respectively. 
	These radio continuum images reveal an extended spherical-like structure, 
	which is related to the $\gamma$ Cygni SNR. 
	Our selected target area is highlighted by a solid box in both the 
	radio continuum images, indicating the location of the NGC 6910 cluster
	 at the border of the SNR. 
	\citet{2002ApJ...571..866U} reported the age of SNR G78.2+2.1 to be 6600 yr. 
	No molecular $^{13}$CO emission is observed toward the NGC 6910 cluster
	\citep[see positions at {\it l} =78.69 deg; {\it b} = 1.96 deg 
	in Figure~8 in][]{2019ApJ...878...54P}. 
	It is likely that the SNR might have influenced its surrounding environments. 
	However, the impact of the young SNR G78.2+2.1 on the formation of the 
	relatively old cluster NGC 6910  ($\sim$4.5 Myr, Section 4.1.4) is unlikely. 
	Figure~\ref{fig1}c displays the NVSS 1.4 GHz continuum map. 
	The NVSS map indicates the presence of the extended diffuse radio continuum emission 
	(1$\sigma$ $\sim$0.45 mJy/beam) showing existence of
	radio continuum clumps toward the cluster, which is not 
	seen in the CGPS continuum images having lower sensitivity 
	(see a solid box in Figures~\ref{fig1}a,~\ref{fig1}b, and~\ref{fig1}c).
	The presence of the radio clumps suggests the existence of embedded
	massive OB stars, implying the ongoing massive star-formation in the region.
	 Using the {\it Planck} image at 850 $\mu$m (or 353 GHz), tracer of the 
	cold dust emission, 
	we show a field hosting the SNR and the cluster NGC 6910 in  Figure~\ref{fig1}d. 
	It seems that the cluster area is not associated with any 
	noticeable cold dust emission.
	
	\begin{figure*}
	\epsscale{1}
	\plotone{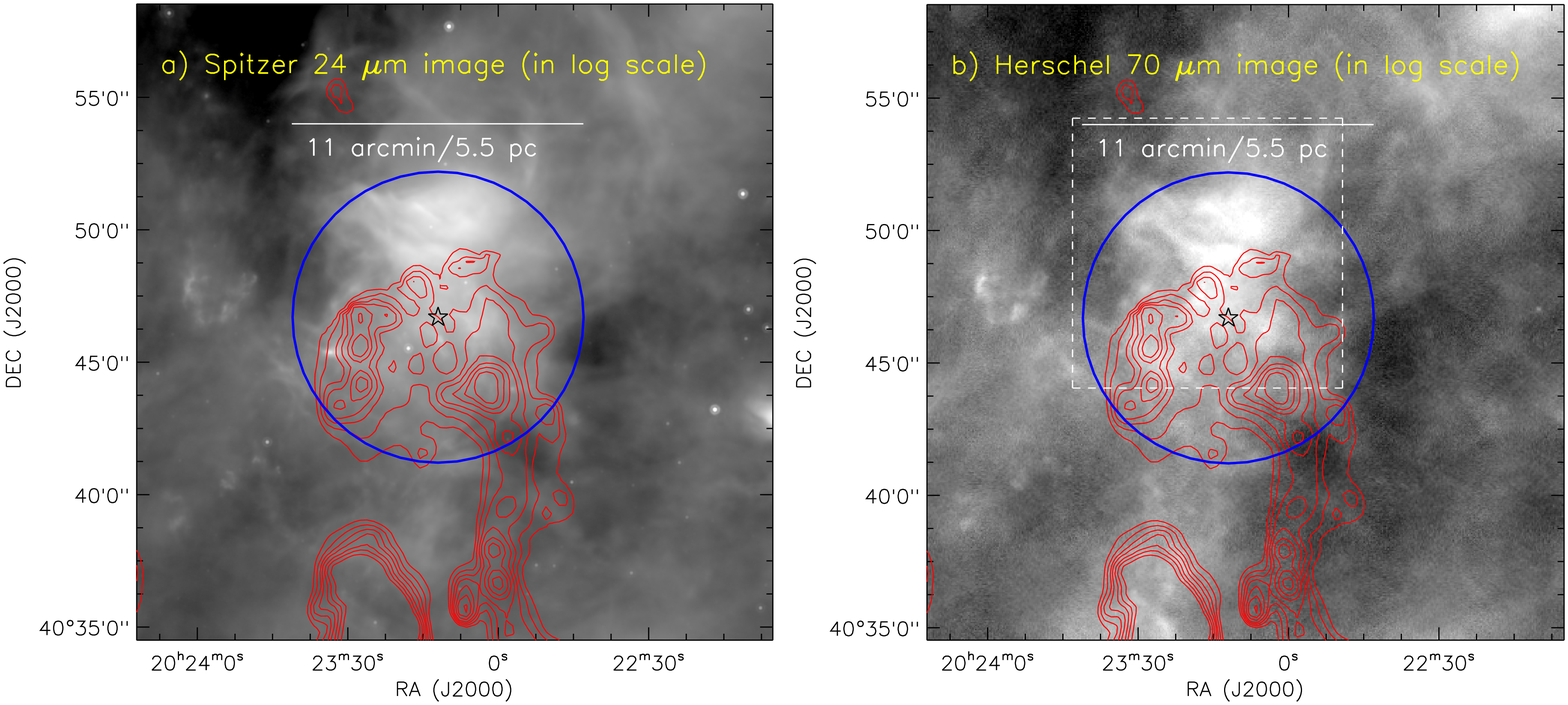}
	\epsscale{1}
	\plotone{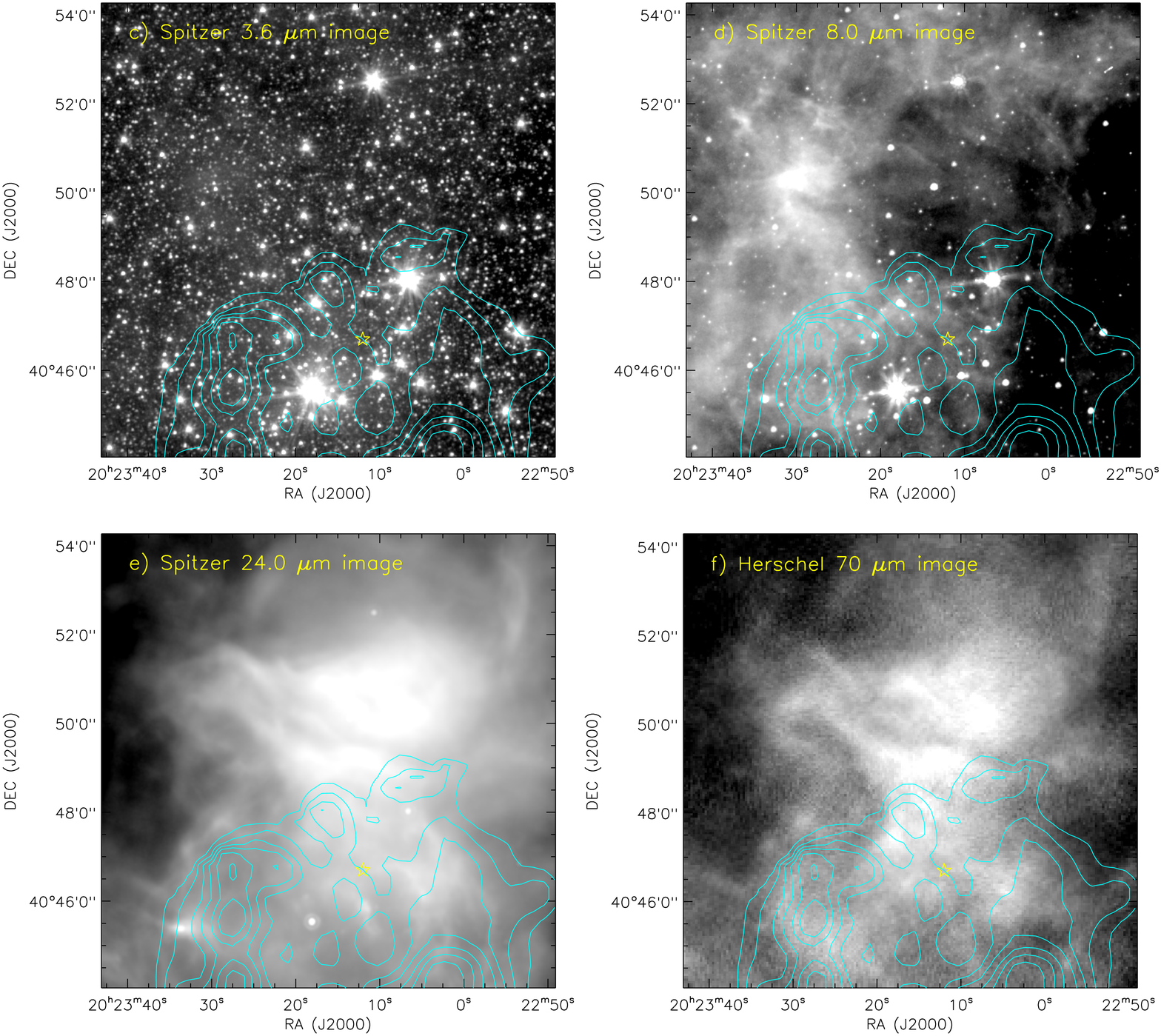}
	\caption{ a) The panel shows the {\it Spitzer} image at 24 $\mu$m overlaid with the NVSS radio continuum
	contours (in red) at 1.4 GHz (see a solid box in Figure~\ref{fig1}a).
	b) Overlay of the 1.4 GHz NVSS radio continuum contours (in red)
	on the {\it Herschel} image at 70 $\mu$m. The white box is shown as a zoomed-in view in Figures~\ref{fig2}c,~\ref{fig2}d,~\ref{fig2}e, and~\ref{fig2}f.
	c) The panel displays the {\it Spitzer} image at 3.6 $\mu$m superimposed with the NVSS radio continuum contours (in cyan) at 1.4 GHz (see a white box in Figure~\ref{fig2}b).
	d) Overlay of the NVSS radio continuum contours (in cyan) at 1.4 GHz on the {\it Spitzer} image at 8.0 $\mu$m.
	e) The panel presents the {\it Spitzer} image at 24 $\mu$m overlaid with the NVSS 1.4 GHz continuum contours (in cyan) at 1.4 GHz.
	f) Overlay of the NVSS 1.4 GHz continuum contours (in cyan) on the {\it Herschel} image at 70 $\mu$m.
	 In the  panels (a) and (b), an extension of the stellar cluster is indicated by a big circle.
	In all the panels, a star symbol represents the central position of the stellar cluster.
	The NVSS contours are shown with the levels of 
	6, 8, 10, 12, 15, 18, and 22 mJy/beam, where 1$\sigma$ $\sim$0.45 mJy/beam.}
	\label{fig2}
	\end{figure*}

	Figures~\ref{fig2}a and~\ref{fig2}b present the {\it Spitzer} 24 $\mu$m 
	and {\it Herschel} 70 $\mu$m images of the area selected in this paper, respectively. 
	Both the infrared images 
	are overlaid with the NVSS 1.4 GHz continuum emission contours. 
	The extension of the stellar cluster is also marked in both the infrared images. 
	The images at 24 $\mu$m and 70 $\mu$m allow to qualitatively 
	trace the warm dust emission 
	present in the cluster NGC 6910. 
	In Figures~\ref{fig2}c,~\ref{fig2}d,~\ref{fig2}e, and~\ref{fig2}f, 
	we present a zoomed-in view of the
	central part of the cluster (see a dashed boxed in Figure 7b) using the 3.6 $\mu$m, 
	8.0 $\mu$m, 24 $\mu$m, and 70 $\mu$m images, respectively. 
	These images are also overlaid with 
	the NVSS 1.4 GHz radio continuum emission contours. 
	Diffuse emission is seen in all these infrared images except at 3.6 $\mu$m. 
	We find that the warm dust emission depicted in 24 $\mu$m and 70 $\mu$m 
	images is surrounded by the {\it Spitzer} 8.0 $\mu$m emission. 
	Note that the {\it Spitzer} band at 8.0 $\mu$m hosts PAH 
	features at 7.7 $\mu$m and 8.6 $\mu$m. 
	Considering the inclusion of PAH features 
	in the 8.0 $\mu$m band, the existence of PDRs is 
	evident in the NGC 6910 region, 
	suggesting the impact of massive OB stars present in the stellar cluster. 

	\begin{figure*}
	\epsscale{1}
	\plotone{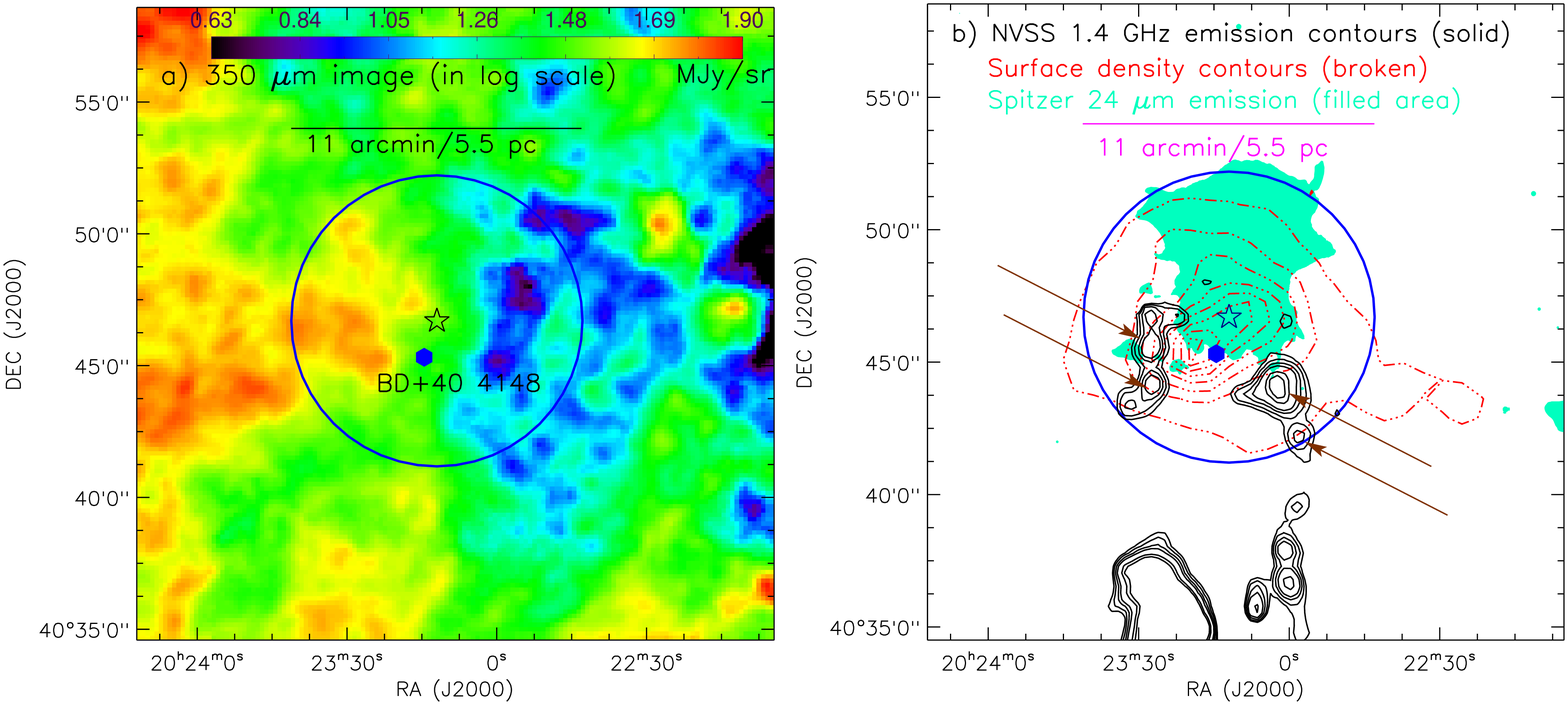}
	\epsscale{1} 
	\plotone{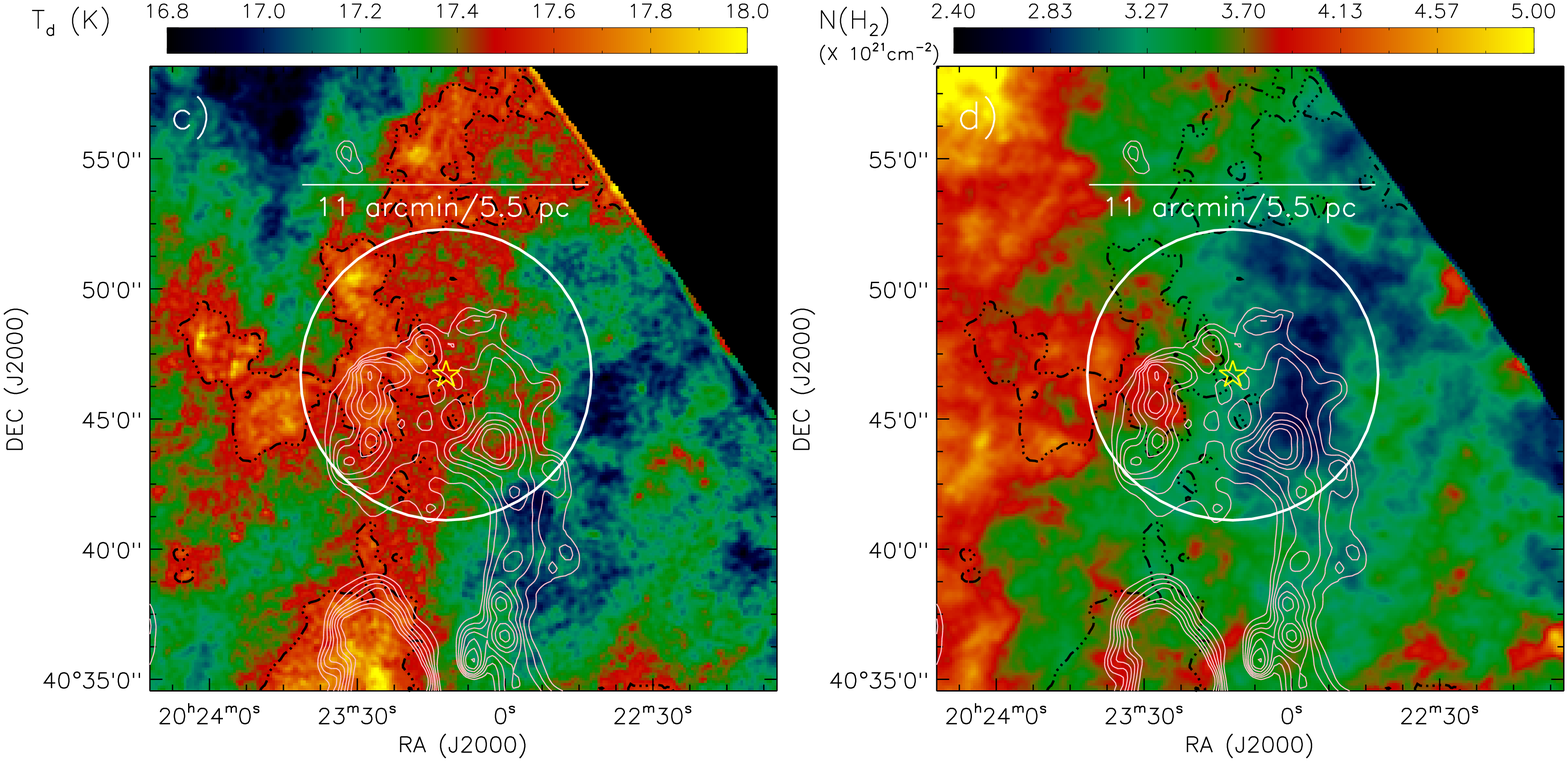}
	\caption{ a) {\it Herschel} image at 350 $\mu$m (see a solid box in 
	Figure~\ref{fig1}a). A filled hexagon indicates the position of a 
	massive star BD+40 4148 
	\citep[spectral class = O9.5V;][]{1965ApJS...12..215H}. 
	b) Overlay of the stellar surface density contours (in red) 
	and the radio continuum emission contours (in black) on the 
	feature traced in the {\it Spitzer} 24 $\mu$m image (i.e., filled area). 	The NVSS contours (in black) are shown with the levels 
	of 11.5, 13, 15, 17, 22, and 25 mJy/beam, where 
	1$\sigma$ $\sim$0.45 mJy/beam. A filled area traced in the {\it Spitzer}
	 24 $\mu$m image is shown with a contour of 58 MJy/sr. The lowest 
	stellar surface density contour (in red) 
	is 1$\sigma$ (i.e., 2.2 stars/pc$^2$ at 1720\ pc), and the step size
	is equal to 1$\sigma$. Arrows highlight the NVSS radio peaks observed
	toward the cluster. 
	c) Overlay of the 1.4 GHz NVSS radio continuum contours (in pink) 
	on the {\it Herschel} temperature map. 
	d) Overlay of the 1.4 GHz NVSS radio continuum contours (in pink) 
	on the  {\it Herschel} column density ($N(\mathrm H_2)$) map. 
	In the panels (c) and (d), the radio emission contours are as same
	as in Figure~\ref{fig2}a, and the  broken contour (in black) indicates
	the dust temperature (T$_{d}$) at 17.5 K. In all the panels, a 
	star symbol represents the central position of the stellar cluster, and an extension of the
	cluster is indicated by a big circle.} 
	\label{fig3}
	\end{figure*}
	
	Figure~\ref{fig3}a shows the sub-millimeter (sub-mm) image at 350 $\mu$m 
	(resolution $\sim$25$''$), 
	where the extension of the stellar cluster is also marked. 
	In the east direction of the cluster, 
	the sub-mm emission is observed, and is not spatially coincident with 
	the 24 $\mu$m and 
	70 $\mu$m emission (see Figures~\ref{fig2}a and~\ref{fig2}b). 
	In Figure~\ref{fig3}b, we have plotted the stellar 
	surface density contours against the distribution of the warm dust emission and the 
	ionized emission, in the cluster area. In Figure~\ref{fig3}b, 
	the warm dust emission is traced using 
	the {\it Spitzer} 24 $\mu$m image (see filled area), 
 	while the NVSS 1.4 GHz continuum 
	emission depicts the ionized emission (see black contours). 
	We find that the radio continuum peaks (ionized clumps or H\,{\sc ii} 
	regions) are located away from the center of the 
	cluster (see arrows in Figure~\ref{fig3}b). 
	Based on the analysis of the NVSS radio continuum data, we calculate that 
	these ionized 
	clumps are powered by 
	B1V--B0.5V stars \citep[see Table II in][for a theoretical value]{1973AJ.....78..929P}, 
	and their dynamical ages are estimated 
	to be $\sim$0.07--0.12 Myr. In this calculation, we have employed the
	same procedures as carried out in \citet{2017ApJ...834...22D}.
	Using the integrated radio continuum flux density and radius (R$_{HII}$) of each 
	ionized clump, the number of Lyman continuum photons (N$_{uv}$) 
	was estimated following 
	the equation given in \citep{1976AJ.....81..172M}. 
	Then, with the knowledge of N$_{uv}$ and R$_{HII}$ values, the age of each ionized 
	clump or H\,{\sc ii} region has been estimated using the 
	equation given in \citet{1980pim..book.....D}.

	The {\it Herschel} temperature and column density maps can be used
        to deduce the physical conditions present in a given star-forming
        region \citep[cf.][]{2019ApJ...884...84D,2019ApJ...877....1D,2019ApJ...878...26D}.
	Figures~\ref{fig3}c and~\ref{fig3}d show the {\it Herschel} temperature and column 
	density maps (resolution $\sim$12$''$), respectively.
	These {\it Herschel} maps are also overlaid with the NVSS radio continuum 
	emission contours. In Figure~\ref{fig3}c, we have traced a feature in the 
	temperature map using a contour of T$_{d}$ = 17.5 K
	 ($\approx average~dust~temperature$). 
	Using a black broken  contour, this feature is also highlighted in 
	both the {\it Herschel} maps. 
	The ionized clumps distributed within the cluster are 
	associated with emission at T$_{d}$ = 17.5--18.0 K. In Figure~\ref{fig3}d, we do not 
	find high column density materials in the direction of the 
	cluster except in the east direction 
	where radio peaks are seen harboring the very young stellar sources.

	\subsubsection{The star formation scenario}

	Our careful analysis of various observational data sets suggests
	that the MF slope of the cluster region is shallower than the Salpeter value
	(i.e., $\Gamma$= -1.35). It shows a signature of the presence of more
	massive stars compared to the low mass stars in the cluster region.
	This argument suggests the effect of mass-segregation.
	A comparison between the cluster age (i.e., $\sim$4.5 Myr) and
	its dynamical relaxation time (i.e. $\sim$6.5 Myr) indicates
	that the cluster is not relaxed yet.
	Furthermore, the observed mass-segregation seen in this cluster may be the
	imprint of their formation processes.
	These properties make the NGC 6910 cluster a special target to study the feedback effect of 
	 massive star(s) on its environment.

	Massive stars can provide positive feedback affecting star-formation by
	accumulating neutral material at the periphery of H\,{\sc ii} regions
	via  `collect and collapse' mechanism \citep{1977ApJ...214..725E, 1994MNRAS.268..291W},
	or by the compression of pre-existing dense condensations 
	via `radiation-driven implosion' mechanism \citep[RDI;][]{1989ApJ...346..735B, 1994A&A...289..559L}.
	Star-formation induced in a region by these processes is also called
	triggered-star-formation and observational signposts usually include age sequence of the 
	stellar sources
	and the distribution of cool, warm and ionized gas in 
	the region \citep[see also,][]
	{2007ApJ...671..555S,2011MNRAS.411.2530J,2007MNRAS.380.1141S,2012PASJ...64..107S,2017MNRAS.467.2943S}.
	Observational studies of bubbles associated with H\,{\sc ii} regions 
	created by massive 
	stars suggest that their expansion probably triggers 
	14\% to 30\%  of star-formation in our Galaxy \citep[e.g.,][]{2010A&A...523A...6D,2012MNRAS.421..408T,2012ApJ...755...71K}, 
	thus implying the importance of massive
	OB stars on the {star-formation} activities in our Galaxy.

	The cluster NGC 6910 contains several massive OB stars and the 
	most massive of them
	is an O9.5V star known as BD+40 4148 \citep{1965ApJS...12..215H}.
	The position of this massive star is marked by a diamond
	in Figure~\ref{color2}.
	In general, a massive star can influence the surroundings through its different feedback
	pressure components (such as, pressure of an H\,{\sc ii} region $(P_{HII})$,
	radiation pressure (P$_{rad}$), and stellar wind ram
	pressure (P$_{wind}$)) \citep[see e.g.,][]{2012ApJ...758L..28B,2017ApJ...834...22D}.
	These pressure components can be expressed as below \citep[see e.g.,][]{2012ApJ...758L..28B}:
        \begin{equation}
        P_{HII} = \mu m_{H} c_{s}^2\, \left(\sqrt{3N_{uv}\over 4\pi\,\alpha_{B}\, D_{s}^3}\right);\\
        \end{equation}
        \begin{equation}
        P_{rad} = L_{bol}/ 4\pi c D_{s}^2; \\
        \end{equation}
        \begin{equation}
        P_{wind} = \dot{M}_{w} V_{w} / 4 \pi D_{s}^2; \\
        \end{equation}
	In the equations above, `N$_{uv}$' is the Lyman continuum photons, `c$_{s}$' is the
	sound speed in the photoionized region \citep[= 11 km s$^{-1}$;][]{2009A&A...497..649B},
	`$\alpha_{B}$' is the radiative recombination coefficient \citep[=  2.6 $\times$ 10$^{-13}$ 
	$\times$ (10$^{4}$ K/T$_{e}$)$^{0.7}$
	cm$^{3}$ s$^{-1}$;][]{1997ApJ...489..284K}, `$\mu$' is the mean molecular weight in
	the ionized gas  \citep[= 0.678;][]{2009A&A...497..649B},
	`m$_{H}$' is the hydrogen atom mass, `$\dot{M}_{w}$' is the mass--loss rate,
	`V$_{w}$' is the wind velocity of the ionizing source, and `L$_{bol}$' is the
	bolometric luminosity of the ionizing source. `D$_{s}$' is the projected distance
	from the location of the O9.5V type star to the ionized clumps,
	which is adopted to be 1.0\ pc (see Figure~\ref{fig3}b).

	We have used $L_{bol}$ = 66070 L$_{\odot}$ \citep[][]{1973AJ.....78..929P}, 
	$\dot{M}_{w}$ $\approx$ 1.58 $\times$ 10$^{-9}$ M$_{\odot}$ yr$^{-1}$ \citep[][]{2009A&A...498..837M}, 		V$_{w}$ $\approx$ 1500 km s$^{-1}$ \citep[][]{2009A&A...498..837M}, and
	N$_{uv}$ = 1.2 $\times$ 10$^{48}$ phs$^{-1}$ \citep[][]{1973AJ.....78..929P}
	for a star of O9.5V spectral type to estimate different pressure 
	components which comes out to be P$_{HII}$ $\approx$ 2.6 $\times$ 10$^{-10}$ dynes\, cm$^{-2}$,
	$P_{rad}$ $\approx$ 7.1 $\times$ 10$^{-11}$ dynes\, cm$^{-2}$, and
	P$_{wind}$ $\approx$ 1.3 $\times$ 10$^{-13}$  dynes\, cm$^{-2}$.
	It gives a total pressure (i.e., P$_{total}$ = P$_{HII}$ + $P_{rad}$ + P$_{wind}$)
	driven by a massive star as $\sim$3.4 $\times$ 10$^{-10}$ dynes\, cm$^{-2}$.
	It appears that the P$_{HII}$ component is relatively higher than other two
	pressure components. Furthermore, we find that the value of $P_{total}$ is also higher than
	the pressure of a typical cool molecular cloud ($P_{MC}$$\sim$10$^{-11}$--10$^{-12}$ 	
	dynes cm$^{-2}$ for a
	temperature $\sim$20 K and particle density $\sim$10$^{3}$--10$^{4}$ cm$^{-3}$) 
	\citep[see Table 7.3 of][]{1980pim..book.....D}.
	It suggests that the massive star seems to have significantly influenced its environment. 
	Additionally, the photoionized gas associated with the cluster appears to be responsible for 
	the feedback mechanism.

	We also find the existence of young ionized clumps,
	located along the edge of the NGC 6910 cluster containing massive stars, in a high column density 
	region.
	The center of the cluster is associated with warm dust emission, and the ionized clumps are 
	distributed in the PDRs.
	Hence, it is likely that these massive stars might have influenced the birth of
	the youngest massive B-type stars (age range $\sim$0.07-0.12 Myr) powering the ionized clumps.
	In a triggered star-forming region `Sh 2-294',
	\citet{2007ApJ...671..555S} have also found ionized clumps away from the exciting source.
	We did not find any ring/arc of gas and dust surrounding the NGC 6910 cluster
	as has been found in the regions showing collect and collapse mechanism
	\citep[for details, cf.][]{2005A&A...433..565D}.
	Therefore, the age difference between the young sources in region
	and the central massive source, and the distribution of PDRs/warm and cold gas and dust/ionized gas,
	seem to be consistent with the hypothesis that the star-formation at the border of NGC 6910 may
	be due to the RDI process. However, with the currently available observations and data, it will be early to
	establish or rule out either of the scenarios.

	\section{Summary and Conclusions}
	\label{sec:conc}

	We have performed deep multiband ($UBV(RI)_c$) and wide-field optical 
	photometric observations ($FOV\sim$ $22\times23$ arcmin$^2$) around the NGC 6910 
	cluster up to 22 mag in the $V$-band. 
	The data is complete down to 20.6 mag in the $V$-band, 
	corresponding to the mass completeness of 0.8 $M_\odot$.
	The optical data along with multiwavelength archival data have been 
	used to study the 
	ongoing physical processes in the cluster NGC 6910.
	The main results are summarized below:

	\begin{itemize}

	\item
	By using the radial density profile of stellar sources, we estimated the 
	cluster radius as 5.5 arcmin with a core radius of 1.4 arcmin.  
	We have used the stellar surface density contours to study the 
	structure of this cluster and found that the stellar surface 
	density contours match
	with the cluster size determined by the RDP. 
	The core region of this cluster seems to be elongated.

	\item
	We have calculated the membership probability for 916 stars in the cluster NGC 6910
	and found 128 member stars with membership 
	probabilities higher than 80\% with $G<20$ mag.
	We have calculated a distance to the cluster as $1.72\pm0.08$ kpc using 
	parallax for cluster members.
        With the help of TCDs and CMDs, we have also estimated
        the foreground reddening $E(B-V)_{min}$, distance, and age of the cluster 
	NGC 6910 to be 0.95 mag, 1.72 kpc, and 4.5 Myr, respectively. 

	\item
	It is found that the mass function slope `$\Gamma$' in the cluster 
	region (i.e. $-0.74\pm0.15$) is shallower than the Salpeter
	value (i.e., -1.35), which indicates the presence of 
	large number of massive stars as compared to low mass stars in the cluster 
	region and indicates an effect of mass-segregation. 
	A contrast between cluster age (i.e., $\sim$4.5 Myr) 
	and its dynamical relaxation time (i.e. $\sim$6.5 Myr) suggests
	that the cluster is not relaxed yet, and the observed mass-segregation 
	seen in this cluster may be the imprint of their formation processes.

	\item
	The distribution of warm dust is found 
	in the central region of the cluster 
	which also contain massive stars. The cluster is surrounded by the PDRs along with the 
	presence of radio peaks (ionized clumps) and cold gas. The total pressure
        (i.e. P$_{total}$=$\sim$3.4 $\times$ 10$^{-10}$ dynes\, cm$^{-2}$) driven by the
        massive O9.5V star (BD+40 4148) at the location of the ionized clumps 
	is also found to be very high as compared to the 
	pressure of a typical cool molecular cloud. 
	All these signatures strongly 
	suggest the influence of massive star(s) of the NGC 6910 cluster
	on its environment.

	\item
        We have determined the spectral type and age of the young star responsible 
	for the radio emission near the border of the cluster (ionized clumps) 
	as B--type and $\sim$0.07--0.12 Myr, respectively.
	The age gradient between the central massive star (4.5 Myr) and the 
	ionized clumps ($\sim$0.07--0.12 Myr) along with other signatures 
	indicates the influence of massive star(s). This suggests that the
	feedback effects from the central massive stars are triggering the formation 
	of next generation of stars in the surrounding region.

	\end{itemize}

	\section*{Acknowledgments}

	 We thank the anonymous reviewer for a critical reading of the 
	manuscript and constructive suggestions, which greatly improved the overall 
	quality of the paper.
	The observations reported in 
	this paper were obtained using the 1-m Sampurnanand Telescope, Nainital, India. 
	This work is based on data obtained as part of the UKIRT Infrared 
	Deep Sky Survey(UKIDSS). 
	This publication made use of data products from the 2MASS (a joint project of the
	University of Massachusetts and the Infrared Processing and 
	Analysis Center/California 
	Institute of Technology, funded by NASA and NSF), archival data 
	obtained with the Spitzer 
	Space Telescope (operated by the Jet Propulsion Laboratory, California Institute of 
	Technology under a contract with NASA). 
	This study has made use of data from the European Space Agency (ESA) 
	mission {\it Gaia} (https://www.cosmos.esa.int/gaia), 
	processed by the {\it Gaia} Data Processing and
	Analysis Consortium (DPAC; https://www.cosmos.esa.int/web/gaia/dpac/consortium).
	Funding for the DPAC has been provided by the institutions participating
	in the {\it Gaia} Multilateral Agreement.  
	 DKO acknowledges the support of the 
	Department of Atomic Energy, Government of India, under 
	project No. 12-R\&D-TFR-5.02-0200. 
	LKD acknowledges the support of the Department of Space, Government of India.

	\software{ESO-MIDAS \citep{1992ASPC...25..120B}, IRAF \citep{1986SPIE..627..733T,1993ASPC...52..173T}, 
	DAOPHOT-II~software \citep{1987PASP...99..191S}}

	\bibliography{n6910}{}

\begin{thebibliography}{}
\expandafter\ifx\csname natexlab\endcsname\relax\def\natexlab#1{#1}\fi
\providecommand{\url}[1]{\href{#1}{#1}}
\providecommand{\dodoi}[1]{doi:~\href{http://doi.org/#1}{\nolinkurl{#1}}}
\providecommand{\doeprint}[1]{\href{http://ascl.net/#1}{\nolinkurl{http://ascl.net/#1}}}
\providecommand{\doarXiv}[1]{\href{https://arxiv.org/abs/#1}{\nolinkurl{https://arxiv.org/abs/#1}}}

\bibitem[{{Adelman} \& {Y{\"u}ce}(2007)}]{2007BaltA..16..311A}
{Adelman}, S.~J., \& {Y{\"u}ce}, K. 2007, Baltic Astronomy, 16, 311

\bibitem[{{Appenzeller} \& {Wendker}(1980)}]{1980A&A....89..239A}
{Appenzeller}, I., \& {Wendker}, H.~J. 1980, \aap, 89, 239

\bibitem[{{Arkhipova} \& {Lozinskaia}(1978)}]{1978AZh....55.1320A}
{Arkhipova}, V.~P., \& {Lozinskaia}, T.~A. 1978, \azh, 55, 1320

\bibitem[{{Baars} \& {Wendker}(1981)}]{1981A&A...101...39B}
{Baars}, J.~W.~M., \& {Wendker}, H.~J. 1981, \aap, 101, 39

\bibitem[{{Bailer-Jones} {et~al.}(2018){Bailer-Jones}, {Rybizki}, {Fouesneau},
  {Mantelet}, \& {Andrae}}]{2018AJ....156...58B}
{Bailer-Jones}, C.~A.~L., {Rybizki}, J., {Fouesneau}, M., {Mantelet}, G., \&
  {Andrae}, R. 2018, \aj, 156, 58, \dodoi{10.3847/1538-3881/aacb21}

\bibitem[{{Balaguer-N{\'u}nez} {et~al.}(1998){Balaguer-N{\'u}nez}, {Tian}, \&
  {Zhao}}]{1998A&AS..133..387B}
{Balaguer-N{\'u}nez}, L., {Tian}, K.~P., \& {Zhao}, J.~L. 1998, \aaps, 133,
  387, \dodoi{10.1051/aas:1998324}

\bibitem[{{Banse} {et~al.}(1992){Banse}, {Grosbol}, \&
  {Baade}}]{1992ASPC...25..120B}
{Banse}, K., {Grosbol}, P., \& {Baade}, D. 1992, Astronomical Society of the
  Pacific Conference Series, Vol.~25, {MIDAS as a Development Environment}, ed.
  D.~M. {Worrall}, C.~{Biemesderfer}, \& J.~{Barnes}, 120

\bibitem[{{Bastian} {et~al.}(2010){Bastian}, {Covey}, \&
  {Meyer}}]{2010ARA&A..48..339B}
{Bastian}, N., {Covey}, K.~R., \& {Meyer}, M.~R. 2010, \araa, 48, 339,
  \dodoi{10.1146/annurev-astro-082708-101642}

\bibitem[{{Battinelli} \& {Capuzzo-Dolcetta}(1991)}]{1991MNRAS.249...76B}
{Battinelli}, P., \& {Capuzzo-Dolcetta}, R. 1991, \mnras, 249, 76,
  \dodoi{10.1093/mnras/249.1.76}

\bibitem[{{Becker} \& {Fenkart}(1971)}]{1971A&AS....4..241B}
{Becker}, W., \& {Fenkart}, R. 1971, \aaps, 4, 241

\bibitem[{{Bellini} {et~al.}(2009){Bellini}, {Piotto}, {Bedin}, {Anderson},
  {Platais}, {Momany}, {Moretti}, {Milone}, \&
  {Ortolani}}]{2009A&A...493..959B}
{Bellini}, A., {Piotto}, G., {Bedin}, L.~R., {et~al.} 2009, \aap, 493, 959,
  \dodoi{10.1051/0004-6361:200810880}

\bibitem[{{Bertoldi}(1989)}]{1989ApJ...346..735B}
{Bertoldi}, F. 1989, \apj, 346, 735, \dodoi{10.1086/168055}

\bibitem[{{Bisbas} {et~al.}(2009){Bisbas}, {W{\"u}nsch}, {Whitworth}, \&
  {Hubber}}]{2009A&A...497..649B}
{Bisbas}, T.~G., {W{\"u}nsch}, R., {Whitworth}, A.~P., \& {Hubber}, D.~A. 2009,
  \aap, 497, 649, \dodoi{10.1051/0004-6361/200811522}

\bibitem[{{Bonnell} {et~al.}(1998){Bonnell}, {Bate}, \&
  {Zinnecker}}]{1998MNRAS.298...93B}
{Bonnell}, I.~A., {Bate}, M.~R., \& {Zinnecker}, H. 1998, \mnras, 298, 93,
  \dodoi{10.1046/j.1365-8711.1998.01590.x}

\bibitem[{{Bossini} {et~al.}(2019){Bossini}, {Vallenari}, {Bragaglia},
  {Cantat-Gaudin}, {Sordo}, {Balaguer-N{\'u}{\~n}ez}, {Jordi}, {Moitinho},
  {Soubiran}, {Casamiquela}, {Carrera}, \& {Heiter}}]{2019A&A...623A.108B}
{Bossini}, D., {Vallenari}, A., {Bragaglia}, A., {et~al.} 2019, \aap, 623,
  A108, \dodoi{10.1051/0004-6361/201834693}

\bibitem[{{Bressert} {et~al.}(2012){Bressert}, {Ginsburg}, {Bally},
  {Battersby}, {Longmore}, \& {Testi}}]{2012ApJ...758L..28B}
{Bressert}, E., {Ginsburg}, A., {Bally}, J., {et~al.} 2012, \apjl, 758, L28,
  \dodoi{10.1088/2041-8205/758/2/L28}

\bibitem[{{Cardelli} {et~al.}(1989){Cardelli}, {Clayton}, \&
  {Mathis}}]{1989ApJ...345..245C}
{Cardelli}, J.~A., {Clayton}, G.~C., \& {Mathis}, J.~S. 1989, \apj, 345, 245,
  \dodoi{10.1086/167900}

\bibitem[{{Carey} {et~al.}(2005){Carey}, {Noriega-Crespo}, {Price}, {Padgett},
  {Kraemer}, {Indebetouw}, {Mizuno}, {Ali}, {Berriman}, \&
  {Boulanger}}]{2005AAS...207.6333C}
{Carey}, S.~J., {Noriega-Crespo}, A., {Price}, S.~D., {et~al.} 2005, in
  American Astronomical Society Meeting Abstracts, Vol. 207, 63.33

\bibitem[{{Chabrier}(2003)}]{2003PASP..115..763C}
{Chabrier}, G. 2003, \pasp, 115, 763, \dodoi{10.1086/376392}

\bibitem[{{Chauhan} {et~al.}(2011){Chauhan}, {Pandey}, {Ogura}, {Jose}, {Ojha},
  {Samal}, \& {Mito}}]{2011MNRAS.415.1202C}
{Chauhan}, N., {Pandey}, A.~K., {Ogura}, K., {et~al.} 2011, \mnras, 415, 1202,
  \dodoi{10.1111/j.1365-2966.2011.18742.x}

\bibitem[{{Chen} {et~al.}(2004){Chen}, {Chen}, \& {Shu}}]{2004AJ....128.2306C}
{Chen}, W.~P., {Chen}, C.~W., \& {Shu}, C.~G. 2004, \aj, 128, 2306,
  \dodoi{10.1086/424855}

\bibitem[{{Chini} {et~al.}(1990){Chini}, {Kruegel}, \&
  {Kreysa}}]{1990A&A...227L...5C}
{Chini}, R., {Kruegel}, E., \& {Kreysa}, E. 1990, \aap, 227, L5

\bibitem[{{Comer{\'o}n} \& {Pasquali}(2012)}]{2012A&A...543A.101C}
{Comer{\'o}n}, F., \& {Pasquali}, A. 2012, \aap, 543, A101,
  \dodoi{10.1051/0004-6361/201219022}

\bibitem[{{Condon} {et~al.}(1998){Condon}, {Cotton}, {Greisen}, {Yin},
  {Perley}, {Taylor}, \& {Broderick}}]{1998AJ....115.1693C}
{Condon}, J.~J., {Cotton}, W.~D., {Greisen}, E.~W., {et~al.} 1998, \aj, 115,
  1693, \dodoi{10.1086/300337}

\bibitem[{{Corbelli} {et~al.}(2005){Corbelli}, {Palla}, \&
  {Zinnecker}}]{2005ASSL..327.....C}
{Corbelli}, E., {Palla}, F., \& {Zinnecker}, H., eds. 2005, Astrophysics and
  Space Science Library, Vol. 327, {The Initial Mass Function 50 years later}

\bibitem[{{Dambis}(1999)}]{1999AstL...25....7D}
{Dambis}, A.~K. 1999, Astronomy Letters, 25, 7

\bibitem[{{Dame} {et~al.}(2001){Dame}, {Hartmann}, \&
  {Thaddeus}}]{2001ApJ...547..792D}
{Dame}, T.~M., {Hartmann}, D., \& {Thaddeus}, P. 2001, \apj, 547, 792,
  \dodoi{10.1086/318388}

\bibitem[{{Davies} \& {Tovmassian}(1963)}]{1963MNRAS.127...45D}
{Davies}, R.~D., \& {Tovmassian}, H.~M. 1963, \mnras, 127, 45,
  \dodoi{10.1093/mnras/127.1.45}

\bibitem[{{Deharveng} {et~al.}(2005){Deharveng}, {Zavagno}, \&
  {Caplan}}]{2005A&A...433..565D}
{Deharveng}, L., {Zavagno}, A., \& {Caplan}, J. 2005, \aap, 433, 565,
  \dodoi{10.1051/0004-6361:20041946}

\bibitem[{{Deharveng} {et~al.}(2010){Deharveng}, {Schuller}, {Anderson},
  {Zavagno}, {Wyrowski}, {Menten}, {Bronfman}, {Testi}, {Walmsley}, \&
  {Wienen}}]{2010A&A...523A...6D}
{Deharveng}, L., {Schuller}, F., {Anderson}, L.~D., {et~al.} 2010, \aap, 523,
  A6, \dodoi{10.1051/0004-6361/201014422}

\bibitem[{{Deharveng} {et~al.}(2015){Deharveng}, {Zavagno}, {Samal},
  {Anderson}, {LeLeu}, {Brevot}, {Duarte-Cabral}, {Molinari}, {Pestalozzi},
  {Foster}, {Rathborne}, \& {Jackson}}]{2015A&A...582A...1D}
{Deharveng}, L., {Zavagno}, A., {Samal}, M.~R., {et~al.} 2015, \aap, 582, A1,
  \dodoi{10.1051/0004-6361/201423835}

\bibitem[{{Delgado} \& {Alfaro}(2000)}]{2000AJ....119.1848D}
{Delgado}, A.~J., \& {Alfaro}, E.~J. 2000, \aj, 119, 1848,
  \dodoi{10.1086/301298}

\bibitem[{{Dewangan}(2019)}]{2019ApJ...884...84D}
{Dewangan}, L.~K. 2019, \apj, 884, 84, \dodoi{10.3847/1538-4357/ab4189}

\bibitem[{{Dewangan} {et~al.}(2016){Dewangan}, {Ojha}, {Luna}, {Anand arao},
  {Ninan}, {Mallick}, \& {Mayya}}]{2016ApJ...819...66D}
{Dewangan}, L.~K., {Ojha}, D.~K., {Luna}, A., {et~al.} 2016, \apj, 819, 66,
  \dodoi{10.3847/0004-637X/819/1/66}

\bibitem[{{Dewangan} {et~al.}(2017){Dewangan}, {Ojha}, {Zinchenko},
  {Janardhan}, \& {Luna}}]{2017ApJ...834...22D}
{Dewangan}, L.~K., {Ojha}, D.~K., {Zinchenko}, I., {Janardhan}, P., \& {Luna},
  A. 2017, \apj, 834, 22, \dodoi{10.3847/1538-4357/834/1/22}

\bibitem[{{Dewangan} {et~al.}(2019{\natexlab{a}}){Dewangan}, {Pirogov},
  {Ryabukhina}, {Ojha}, \& {Zinchenko}}]{2019ApJ...877....1D}
{Dewangan}, L.~K., {Pirogov}, L.~E., {Ryabukhina}, O.~L., {Ojha}, D.~K., \&
  {Zinchenko}, I. 2019{\natexlab{a}}, \apj, 877, 1,
  \dodoi{10.3847/1538-4357/ab1aa6}

\bibitem[{{Dewangan} {et~al.}(2019{\natexlab{b}}){Dewangan}, {Sano}, {Enokiya},
  {Tachihara}, {Fukui}, \& {Ojha}}]{2019ApJ...878...26D}
{Dewangan}, L.~K., {Sano}, H., {Enokiya}, R., {et~al.} 2019{\natexlab{b}},
  \apj, 878, 26, \dodoi{10.3847/1538-4357/ab1cba}

\bibitem[{{Dickel} {et~al.}(1977){Dickel}, {Seacord}, \&
  {Gottesman}}]{1977ApJ...218..133D}
{Dickel}, H.~R., {Seacord}, II, A.~W., \& {Gottesman}, S.~T. 1977, \apj, 218,
  133, \dodoi{10.1086/155665}

\bibitem[{{Dyson} \& {Williams}(1980)}]{1980pim..book.....D}
{Dyson}, J.~E., \& {Williams}, D.~A. 1980, {Physics of the interstellar medium}

\bibitem[{{Elmegreen} \& {Lada}(1977)}]{1977ApJ...214..725E}
{Elmegreen}, B.~G., \& {Lada}, C.~J. 1977, \apj, 214, 725,
  \dodoi{10.1086/155302}

\bibitem[{{Friel} {et~al.}(2014){Friel}, {Donati}, {Bragaglia}, {Jacobson},
  {Magrini}, {Prisinzano}, {Rand ich}, {Tosi}, {Cantat-Gaudin}, {Vallenari},
  {Smiljanic}, {Carraro}, {Sordo}, {Maiorca}, {Tautvai{\v{s}}ien{\.{e}}},
  {Sestito}, {Zaggia}, {Jim{\'e}nez-Esteban}, {Gilmore}, {Jeffries}, {Alfaro},
  {Bensby}, {Koposov}, {Korn}, {Pancino}, {Recio-Blanco}, {Franciosini},
  {Hill}, {Jackson}, {de Laverny}, {Morbidelli}, {Sacco}, {Worley},
  {Hourihane}, {Costado}, {Jofr{\'e}}, \& {Lind}}]{2014A&A...563A.117F}
{Friel}, E.~D., {Donati}, P., {Bragaglia}, A., {et~al.} 2014, \aap, 563, A117,
  \dodoi{10.1051/0004-6361/201323215}

\bibitem[{{Gaia Collaboration} {et~al.}(2018{\natexlab{a}}){Gaia
  Collaboration}, {Katz}, {Antoja}, {Romero-G{\'o}mez}, {Drimmel}, {Reyl{\'e}},
  {Seabroke}, {Soubiran}, {Babusiaux}, {Di Matteo}, {Figueras}, {Poggio},
  {Robin}, {Evans}, {Brown}, {Vallenari}, {Prusti}, {de Bruijne},
  {Bailer-Jones}, {Biermann}, {Eyer}, {Jansen}, {Jordi}, {Klioner}, {Lammers},
  {Lindegren}, {Luri}, {Mignard}, {Panem}, {Pourbaix}, {Randich}, {Sartoretti},
  {Siddiqui}, {van Leeuwen}, {Walton}, {Arenou}, {Bastian}, {Cropper},
  {Lattanzi}, {Bakker}, {Cacciari}, {Casta n}, {Chaoul}, {Cheek}, {De Angeli},
  {Fabricius}, {Guerra}, {Holl}, {Masana}, {Messineo}, {Mowlavi},
  {Nienartowicz}, {Panuzzo}, {Portell}, {Riello}, {Tanga}, {Th{\'e}venin},
  {Gracia-Abril}, {Comoretto}, {Garcia-Reinaldos}, {Teyssier}, {Altmann},
  {Andrae}, {Audard}, {Bellas-Velidis}, {Benson}, {Berthier}, {Blomme},
  {Burgess}, {Busso}, {Carry}, {Cellino}, {Clementini}, {Clotet}, {Creevey},
  {Davidson}, {De Ridder}, {Delchambre}, {Dell'Oro}, {Ducourant},
  {Fern{\'a}ndez-Hern{\'a}ndez}, {Fouesneau}, {Fr{\'e}mat}, {Galluccio},
  {Garc{\'\i}a-Torres}, {Gonz{\'a}lez-N{\'u}{\~n}ez}, {Gonz{\'a}lez-Vidal},
  {Gosset}, {Guy}, {Halbwachs}, {Hambly}, {Harrison}, {Hern{\'a}ndez},
  {Hestroffer}, {Hodgkin}, {Hutton}, {Jasniewicz}, {Jean-Antoine-Piccolo},
  {Jordan}, {Korn}, {Krone-Martins}, {Lanzafame}, {Lebzelter}, {L{\"o}ffler},
  {Manteiga}, {Marrese}, {Mart{\'\i}n-Fleitas}, {Moitinho}, {Mora}, {Muinonen},
  {Osinde}, {Pancino}, {Pauwels}, {Petit}, {Recio-Blanco}, {Richards},
  {Rimoldini}, {Sarro}, {Siopis}, {Smith}, {Sozzetti}, {S{\"u}veges}, {Torra},
  {van Reeven}, {Abbas}, {Abreu Aramburu}, {Accart}, {Aerts}, {Altavilla},
  {{\'A}lvarez}, {Alvarez}, {Alves}, {Anderson}, {Andrei}, {Anglada Varela},
  {Antiche}, {Arcay}, {Astraatmadja}, {Bach}, {Baker},
  {Balaguer-N{\'u}{\~n}ez}, {Balm}, {Barache}, {Barata}, {Barbato}, {Barblan},
  {Barklem}, {Barrado}, {Barros}, {Barstow}, {Bartholom{\'e} Mu{\~n}oz},
  {Bassilana}, {Becciani}, {Bellazzini}, {Berihuete}, {Bertone}, {Bianchi},
  {Bienaym{\'e}}, {Blanco-Cuaresma}, {Boch}, {Boeche}, {Bombrun}, {Borrachero},
  {Bossini}, {Bouquillon}, {Bourda}, {Bragaglia}, {Bramante}, {Breddels},
  {Bressan}, {Brouillet}, {Br{\"u}semeister}, {Brugaletta}, {Bucciarelli},
  {Burlacu}, {Busonero}, {Butkevich}, {Buzzi}, {Caffau}, {Cancelliere},
  {Cannizzaro}, {Cantat-Gaudin}, {Carballo}, {Carlucci}, {Carrasco},
  {Casamiquela}, {Castellani}, {Castro-Ginard}, {Charlot}, {Chemin},
  {Chiavassa}, {Cocozza}, {Costigan}, {Cowell}, {Crifo}, {Crosta}, {Crowley},
  {Cuypers}, {Dafonte}, {Damerdji}, {Dapergolas}, {David}, {David}, {de
  Laverny}, {De Luise}, {De March}, {de Souza}, {de Torres}, {Debosscher}, {del
  Pozo}, {Delbo}, {Delgado}, {Delgado}, {Diakite}, {Diener}, {Distefano},
  {Dolding}, {Drazinos}, {Dur{\'a}n}, {Edvardsson}, {Enke}, {Eriksson},
  {Esquej}, {Eynard Bontemps}, {Fabre}, {Fabrizio}, {Faigler}, {Falc a},
  {Farr{\`a}s Casas}, {Federici}, {Fedorets}, {Fernique}, {Filippi},
  {Findeisen}, {Fonti}, {Fraile}, {Fraser}, {Fr{\'e}zouls}, {Gai}, {Galleti},
  {Garabato}, {Garc{\'\i}a-Sedano}, {Garofalo}, {Garralda}, {Gavel}, {Gavras},
  {Gerssen}, {Geyer}, {Giacobbe}, {Gilmore}, {Girona}, {Giuffrida}, {Glass},
  {Gomes}, {Granvik}, {Gueguen}, {Guerrier}, {Guiraud}, {Guti{\'e}}, {Haigron},
  {Hatzidimitriou}, {Hauser}, {Haywood}, {Heiter}, {Helmi}, {Heu}, {Hilger},
  {Hobbs}, {Hofmann}, {Holland }, {Huckle}, {Hypki}, {Icardi}, {Jan{\ss}en},
  {Jevardat de Fombelle}, {Jonker}, {Juh{\'a}sz}, {Julbe}, {Karampelas},
  {Kewley}, {Klar}, {Kochoska}, {Kohley}, {Kolenberg}, {Kontizas}, {Kontizas},
  {Koposov}, {Kordopatis}, {Kostrzewa-Rutkowska}, {Koubsky}, {Lambert},
  {Lanza}, {Lasne}, {Lavigne}, {Le Fustec}, {Le Poncin-Lafitte}, {Lebreton},
  {Leccia}, {Leclerc}, {Lecoeur-Taibi}, {Lenhardt}, {Leroux}, {Liao}, {Licata},
  {Lindstr{\o}m}, {Lister}, {Livanou}, {Lobel}, {L{\'o}pez}, {Managau}, {Mann},
  {Mantelet}, {Marchal}, {Marchant}, {Marconi}, {Marinoni}, {Marschalk{\'o}},
  {Marshall}, {Martino}, {Marton}, {Mary}, {Massari}, {Matijevi{\v{c}}},
  {Mazeh}, {McMillan}, {Messina}, {Michalik}, {Millar}, {Molina}, {Molinaro},
  {Moln{\'a}r}, {Montegriffo}, {Mor}, {Morbidelli}, {Morel}, {Morris},
  {Mulone}, {Muraveva}, {Musella}, {Nelemans}, {Nicastro}, {Noval},
  {O'Mullane}, {Ord{\'e}novic}, {Ord{\'o}{\~n}ez-Blanco}, {Osborne}, {Pagani},
  {Pagano}, {Pailler}, {Palacin}, {Palaversa}, {Panahi}, {Pawlak},
  {Piersimoni}, {Pineau}, {Plachy}, {Plum}, {Poujoulet}, {Pr{\v{s}}a},
  {Pulone}, {Racero}, {Ragaini}, {Rambaux}, {Ramos-Lerate}, {Regibo}, {Riclet},
  {Ripepi}, {Riva}, {Rivard}, {Rixon}, {Roegiers}, {Roelens}, {Rowell},
  {Royer}, {Ruiz-Dern}, {Sadowski}, {Sagrist{\`a} Sell{\'e}s}, {Sahlmann},
  {Salgado}, {Salguero}, {Sanna}, {Santana-Ros}, {Sarasso}, {Savietto},
  {Schultheis}, {Sciacca}, {Segol}, {Segovia}, {S{\'e}gransan}, {Shih},
  {Siltala}, {Silva}, {Smart}, {Smith}, {Solano}, {Solitro}, {Sordo}, {Soria
  Nieto}, {Souchay}, {Spagna}, {Spoto}, {Stampa}, {Steele},
  {Steidelm{\"u}ller}, {Stephenson}, {Stoev}, {Suess}, {Surdej}, {Szabados},
  {Szegedi-Elek}, {Tapiador}, {Taris}, {Tauran}, {Taylor}, {Teixeira},
  {Terrett}, {Teyssand ier}, {Thuillot}, {Titarenko}, {Torra Clotet}, {Turon},
  {Ulla}, {Utrilla}, {Uzzi}, {Vaillant}, {Valentini}, {Valette}, {van Elteren},
  {Van Hemelryck}, {van Leeuwen}, {Vaschetto}, {Vecchiato}, {Veljanoski},
  {Viala}, {Vicente}, {Vogt}, {von Essen}, {Voss}, {Votruba}, {Voutsinas},
  {Walmsley}, {Weiler}, {Wertz}, {Wevers}, {Wyrzykowski}, {Yoldas},
  {{\v{Z}}erjal}, {Ziaeepour}, {Zorec}, {Zschocke}, {Zucker}, {Zurbach}, \&
  {Zwitter}}]{gaia1234}
{Gaia Collaboration}, {Katz}, D., {Antoja}, T., {et~al.} 2018{\natexlab{a}},
  \aap, 616, A11, \dodoi{10.1051/0004-6361/201832865}

\bibitem[{{Gaia Collaboration} {et~al.}(2018{\natexlab{b}}){Gaia
  Collaboration}, {Brown}, {Vallenari}, {Prusti}, {de Bruijne}, {Babusiaux},
  {Bailer-Jones}, {Biermann}, {Evans}, {Eyer}, {Jansen}, {Jordi}, {Klioner},
  {Lammers}, {Lindegren}, {Luri}, {Mignard}, {Panem}, {Pourbaix}, {Randich},
  {Sartoretti}, {Siddiqui}, {Soubiran}, {van Leeuwen}, {Walton}, {Arenou},
  {Bastian}, {Cropper}, {Drimmel}, {Katz}, {Lattanzi}, {Bakker}, {Cacciari},
  {Casta{\~n}eda}, {Chaoul}, {Cheek}, {De Angeli}, {Fabricius}, {Guerra},
  {Holl}, {Masana}, {Messineo}, {Mowlavi}, {Nienartowicz}, {Panuzzo},
  {Portell}, {Riello}, {Seabroke}, {Tanga}, {Th{\'e}venin}, {Gracia-Abril},
  {Comoretto}, {Garcia-Reinaldos}, {Teyssier}, {Altmann}, {Andrae}, {Audard},
  {Bellas-Velidis}, {Benson}, {Berthier}, {Blomme}, {Burgess}, {Busso},
  {Carry}, {Cellino}, {Clementini}, {Clotet}, {Creevey}, {Davidson}, {De
  Ridder}, {Delchambre}, {Dell'Oro}, {Ducourant},
  {Fern{\'a}ndez-Hern{\'a}ndez}, {Fouesneau}, {Fr{\'e}mat}, {Galluccio},
  {Garc{\'\i}a-Torres}, {Gonz{\'a}lez-N{\'u}{\~n}ez}, {Gonz{\'a}lez-Vidal},
  {Gosset}, {Guy}, {Halbwachs}, {Hambly}, {Harrison}, {Hern{\'a}ndez},
  {Hestroffer}, {Hodgkin}, {Hutton}, {Jasniewicz}, {Jean-Antoine-Piccolo},
  {Jordan}, {Korn}, {Krone-Martins}, {Lanzafame}, {Lebzelter}, {L{\"o}ffler},
  {Manteiga}, {Marrese}, {Mart{\'\i}n-Fleitas}, {Moitinho}, {Mora}, {Muinonen},
  {Osinde}, {Pancino}, {Pauwels}, {Petit}, {Recio-Blanco}, {Richards},
  {Rimoldini}, {Robin}, {Sarro}, {Siopis}, {Smith}, {Sozzetti}, {S{\"u}veges},
  {Torra}, {van Reeven}, {Abbas}, {Abreu Aramburu}, {Accart}, {Aerts},
  {Altavilla}, {{\'A}lvarez}, {Alvarez}, {Alves}, {Anderson}, {Andrei},
  {Anglada Varela}, {Antiche}, {Antoja}, {Arcay}, {Astraatmadja}, {Bach},
  {Baker}, {Balaguer-N{\'u}{\~n}ez}, {Balm}, {Barache}, {Barata}, {Barbato},
  {Barblan}, {Barklem}, {Barrado}, {Barros}, {Barstow}, {Bartholom{\'e}
  Mu{\~n}oz}, {Bassilana}, {Becciani}, {Bellazzini}, {Berihuete}, {Bertone},
  {Bianchi}, {Bienaym{\'e}}, {Blanco-Cuaresma}, {Boch}, {Boeche}, {Bombrun},
  {Borrachero}, {Bossini}, {Bouquillon}, {Bourda}, {Bragaglia}, {Bramante},
  {Breddels}, {Bressan}, {Brouillet}, {Br{\"u}semeister}, {Brugaletta},
  {Bucciarelli}, {Burlacu}, {Busonero}, {Butkevich}, {Buzzi}, {Caffau},
  {Cancelliere}, {Cannizzaro}, {Cantat-Gaudin}, {Carballo}, {Carlucci},
  {Carrasco}, {Casamiquela}, {Castellani}, {Castro-Ginard}, {Charlot},
  {Chemin}, {Chiavassa}, {Cocozza}, {Costigan}, {Cowell}, {Crifo}, {Crosta},
  {Crowley}, {Cuypers}, {Dafonte}, {Damerdji}, {Dapergolas}, {David}, {David},
  {de Laverny}, {De Luise}, {De March}, {de Martino}, {de Souza}, {de Torres},
  {Debosscher}, {del Pozo}, {Delbo}, {Delgado}, {Delgado}, {Di Matteo},
  {Diakite}, {Diener}, {Distefano}, {Dolding}, {Drazinos}, {Dur{\'a}n},
  {Edvardsson}, {Enke}, {Eriksson}, {Esquej}, {Eynard Bontemps}, {Fabre},
  {Fabrizio}, {Faigler}, {Falc{\~a}o}, {Farr{\`a}s Casas}, {Federici},
  {Fedorets}, {Fernique}, {Figueras}, {Filippi}, {Findeisen}, {Fonti},
  {Fraile}, {Fraser}, {Fr{\'e}zouls}, {Gai}, {Galleti}, {Garabato},
  {Garc{\'\i}a-Sedano}, {Garofalo}, {Garralda}, {Gavel}, {Gavras}, {Gerssen},
  {Geyer}, {Giacobbe}, {Gilmore}, {Girona}, {Giuffrida}, {Glass}, {Gomes},
  {Granvik}, {Gueguen}, {Guerrier}, {Guiraud}, {Guti{\'e}rrez-S{\'a}nchez},
  {Haigron}, {Hatzidimitriou}, {Hauser}, {Haywood}, {Heiter}, {Helmi}, {Heu},
  {Hilger}, {Hobbs}, {Hofmann}, {Holland}, {Huckle}, {Hypki}, {Icardi},
  {Jan{\ss}en}, {Jevardat de Fombelle}, {Jonker}, {Juh{\'a}sz}, {Julbe},
  {Karampelas}, {Kewley}, {Klar}, {Kochoska}, {Kohley}, {Kolenberg},
  {Kontizas}, {Kontizas}, {Koposov}, {Kordopatis}, {Kostrzewa-Rutkowska},
  {Koubsky}, {Lambert}, {Lanza}, {Lasne}, {Lavigne}, {Le Fustec}, {Le
  Poncin-Lafitte}, {Lebreton}, {Leccia}, {Leclerc}, {Lecoeur-Taibi},
  {Lenhardt}, {Leroux}, {Liao}, {Licata}, {Lindstr{\o}m}, {Lister}, {Livanou},
  {Lobel}, {L{\'o}pez}, {Managau}, {Mann}, {Mantelet}, {Marchal}, {Marchant},
  {Marconi}, {Marinoni}, {Marschalk{\'o}}, {Marshall}, {Martino}, {Marton},
  {Mary}, {Massari}, {Matijevi{\v{c}}}, {Mazeh}, {McMillan}, {Messina},
  {Michalik}, {Millar}, {Molina}, {Molinaro}, {Moln{\'a}r}, {Montegriffo},
  {Mor}, {Morbidelli}, {Morel}, {Morris}, {Mulone}, {Muraveva}, {Musella},
  {Nelemans}, {Nicastro}, {Noval}, {O'Mullane}, {Ord{\'e}novic},
  {Ord{\'o}{\~n}ez-Blanco}, {Osborne}, {Pagani}, {Pagano}, {Pailler},
  {Palacin}, {Palaversa}, {Panahi}, {Pawlak}, {Piersimoni}, {Pineau}, {Plachy},
  {Plum}, {Poggio}, {Poujoulet}, {Pr{\v{s}}a}, {Pulone}, {Racero}, {Ragaini},
  {Rambaux}, {Ramos-Lerate}, {Regibo}, {Reyl{\'e}}, {Riclet}, {Ripepi}, {Riva},
  {Rivard}, {Rixon}, {Roegiers}, {Roelens}, {Romero-G{\'o}mez}, {Rowell},
  {Royer}, {Ruiz-Dern}, {Sadowski}, {Sagrist{\`a} Sell{\'e}s}, {Sahlmann},
  {Salgado}, {Salguero}, {Sanna}, {Santana-Ros}, {Sarasso}, {Savietto},
  {Schultheis}, {Sciacca}, {Segol}, {Segovia}, {S{\'e}gransan}, {Shih},
  {Siltala}, {Silva}, {Smart}, {Smith}, {Solano}, {Solitro}, {Sordo}, {Soria
  Nieto}, {Souchay}, {Spagna}, {Spoto}, {Stampa}, {Steele},
  {Steidelm{\"u}ller}, {Stephenson}, {Stoev}, {Suess}, {Surdej}, {Szabados},
  {Szegedi-Elek}, {Tapiador}, {Taris}, {Tauran}, {Taylor}, {Teixeira},
  {Terrett}, {Teyssand ier}, {Thuillot}, {Titarenko}, {Torra Clotet}, {Turon},
  {Ulla}, {Utrilla}, {Uzzi}, {Vaillant}, {Valentini}, {Valette}, {van Elteren},
  {Van Hemelryck}, {van Leeuwen}, {Vaschetto}, {Vecchiato}, {Veljanoski},
  {Viala}, {Vicente}, {Vogt}, {von Essen}, {Voss}, {Votruba}, {Voutsinas},
  {Walmsley}, {Weiler}, {Wertz}, {Wevers}, {Wyrzykowski}, {Yoldas},
  {{\v{Z}}erjal}, {Ziaeepour}, {Zorec}, {Zschocke}, {Zucker}, {Zurbach}, \&
  {Zwitter}}]{gaia5678}
{Gaia Collaboration}, {Brown}, A.~G.~A., {Vallenari}, A., {et~al.}
  2018{\natexlab{b}}, \aap, 616, A1, \dodoi{10.1051/0004-6361/201833051}

\bibitem[{{Girard} {et~al.}(1989){Girard}, {Grundy}, {Lopez}, \& {van
  Altena}}]{1989AJ.....98..227G}
{Girard}, T.~M., {Grundy}, W.~M., {Lopez}, C.~E., \& {van Altena}, W.~F. 1989,
  \aj, 98, 227, \dodoi{10.1086/115139}

\bibitem[{{Golay}(1974)}]{1974ASSL...41.....G}
{Golay}, M., ed. 1974, Astrophysics and Space Science Library, Vol.~41,
  {Introduction to astronomical photometry}

\bibitem[{{Guetter} \& {Vrba}(1989)}]{1989AJ.....98..611G}
{Guetter}, H.~H., \& {Vrba}, F.~J. 1989, \aj, 98, 611, \dodoi{10.1086/115161}

\bibitem[{{Gutermuth} {et~al.}(2005){Gutermuth}, {Megeath}, {Pipher},
  {Williams}, {Allen}, {Myers}, \& {Raines}}]{2005ApJ...632..397G}
{Gutermuth}, R.~A., {Megeath}, S.~T., {Pipher}, J.~L., {et~al.} 2005, \apj,
  632, 397, \dodoi{10.1086/432460}

\bibitem[{{Harris}(1976)}]{1976ApJS...30..451H}
{Harris}, G.~L.~H. 1976, \apjs, 30, 451, \dodoi{10.1086/190368}

\bibitem[{{Hoag} \& {Applequist}(1965)}]{1965ApJS...12..215H}
{Hoag}, A.~A., \& {Applequist}, N.~L. 1965, \apjs, 12, 215,
  \dodoi{10.1086/190125}

\bibitem[{{Humphreys}(1978)}]{1978ApJS...38..309H}
{Humphreys}, R.~M. 1978, \apjs, 38, 309, \dodoi{10.1086/190559}

\bibitem[{{Hur} {et~al.}(2012){Hur}, {Sung}, \&
  {Bessell}}]{2012AJ....143...41H}
{Hur}, H., {Sung}, H., \& {Bessell}, M.~S. 2012, \aj, 143, 41,
  \dodoi{10.1088/0004-6256/143/2/41}

\bibitem[{{Jose} {et~al.}(2017){Jose}, {Herczeg}, {Samal}, {Fang}, \&
  {Panwar}}]{2017ApJ...836...98J}
{Jose}, J., {Herczeg}, G.~J., {Samal}, M.~R., {Fang}, Q., \& {Panwar}, N. 2017,
  \apj, 836, 98, \dodoi{10.3847/1538-4357/836/1/98}

\bibitem[{{Jose} {et~al.}(2011){Jose}, {Pandey}, {Ogura}, {Ojha}, {Bhatt},
  {Samal}, {Chauhan}, {Sahu}, \& {Rawat}}]{2011MNRAS.411.2530J}
{Jose}, J., {Pandey}, A.~K., {Ogura}, K., {et~al.} 2011, \mnras, 411, 2530,
  \dodoi{10.1111/j.1365-2966.2010.17860.x}

\bibitem[{{Jose} {et~al.}(2013){Jose}, {Pandey}, {Samal}, {Ojha}, {Ogura},
  {Kim}, {Kobayashi}, {Goyal}, {Chauhan}, \& {Eswaraiah}}]{2013MNRAS.432.3445J}
{Jose}, J., {Pandey}, A.~K., {Samal}, M.~R., {et~al.} 2013, \mnras, 432, 3445,
  \dodoi{10.1093/mnras/stt700}

\bibitem[{{Kaluzny} \& {Udalski}(1992)}]{1992AcA....42...29K}
{Kaluzny}, J., \& {Udalski}, A. 1992, \actaa, 42, 29

\bibitem[{{Kendrew} {et~al.}(2012){Kendrew}, {Simpson}, {Bressert}, {Povich},
  {Sherman}, {Lintott}, {Robitaille}, {Schawinski}, \&
  {Wolf-Chase}}]{2012ApJ...755...71K}
{Kendrew}, S., {Simpson}, R., {Bressert}, E., {et~al.} 2012, \apj, 755, 71,
  \dodoi{10.1088/0004-637X/755/1/71}

\bibitem[{{King}(1962)}]{1962AJ.....67..471K}
{King}, I. 1962, \aj, 67, 471, \dodoi{10.1086/108756}

\bibitem[{{Kolaczkowski} {et~al.}(2004){Kolaczkowski}, {Pigulski}, {Kopacki},
  \& {Michalska}}]{kola1234}
{Kolaczkowski}, Z., {Pigulski}, A., {Kopacki}, G., \& {Michalska}, G. 2004,
  \actaa, 54, 33

\bibitem[{{Kroupa}(2002)}]{2002Sci...295...82K}
{Kroupa}, P. 2002, Science, 295, 82, \dodoi{10.1126/science.1067524}

\bibitem[{{Kub{\'a}t} {et~al.}(2007){Kub{\'a}t}, {Kor{\v c}{\'a}kov{\'a}},
  {Kawka}, {Pigulski}, {{\v S}lechta}, \& {{\v S}koda}}]{2007A&A...472..163K}
{Kub{\'a}t}, J., {Kor{\v c}{\'a}kov{\'a}}, D., {Kawka}, A., {et~al.} 2007,
  \aap, 472, 163, \dodoi{10.1051/0004-6361:20077171}

\bibitem[{{Kumar} {et~al.}(2014){Kumar}, {Sharma}, {Manfroid}, {Gosset},
  {Rauw}, {Naz{\'e}}, \& {Kesh Yadav}}]{2014A&A...567A.109K}
{Kumar}, B., {Sharma}, S., {Manfroid}, J., {et~al.} 2014, \aap, 567, A109,
  \dodoi{10.1051/0004-6361/201323027}

\bibitem[{{Kwan}(1997)}]{1997ApJ...489..284K}
{Kwan}, J. 1997, \apj, 489, 284, \dodoi{10.1086/304773}

\bibitem[{{Landecker} {et~al.}(1980){Landecker}, {Roger}, \&
  {Higgs}}]{1980A&AS...39..133L}
{Landecker}, T.~L., {Roger}, R.~S., \& {Higgs}, L.~A. 1980, \aaps, 39, 133

\bibitem[{{Landolt}(1992)}]{1992AJ....104..340L}
{Landolt}, A.~U. 1992, \aj, 104, 340, \dodoi{10.1086/116242}

\bibitem[{{Lawrence} {et~al.}(2007){Lawrence}, {Warren}, {Almaini}, {Edge},
  {Hambly}, {Jameson}, {Lucas}, {Casali}, {Adamson}, {Dye}, {Emerson},
  {Foucaud}, {Hewett}, {Hirst}, {Hodgkin}, {Irwin}, {Lodieu}, {McMahon},
  {Simpson}, {Smail}, {Mortlock}, \& {Folger}}]{2007MNRAS.379.1599L}
{Lawrence}, A., {Warren}, S.~J., {Almaini}, O., {et~al.} 2007, \mnras, 379,
  1599, \dodoi{10.1111/j.1365-2966.2007.12040.x}

\bibitem[{{Lefloch} \& {Lazareff}(1994)}]{1994A&A...289..559L}
{Lefloch}, B., \& {Lazareff}, B. 1994, \aap, 289, 559

\bibitem[{{Lim} {et~al.}(2011){Lim}, {Sung}, {Karimov}, \&
  {Ibrahimov}}]{2011JKAS...44...39L}
{Lim}, B., {Sung}, H.~S., {Karimov}, R., \& {Ibrahimov}, M. 2011, Journal of
  Korean Astronomical Society, 44, 39.
\newblock \doarXiv{1103.4927}

\bibitem[{{Marcolino} {et~al.}(2009){Marcolino}, {Bouret}, {Martins},
  {Hillier}, {Lanz}, \& {Escolano}}]{2009A&A...498..837M}
{Marcolino}, W.~L.~F., {Bouret}, J.~C., {Martins}, F., {et~al.} 2009, \aap,
  498, 837, \dodoi{10.1051/0004-6361/200811289}

\bibitem[{{Marsh} {et~al.}(2015){Marsh}, {Whitworth}, \&
  {Lomax}}]{2015MNRAS.454.4282M}
{Marsh}, K.~A., {Whitworth}, A.~P., \& {Lomax}, O. 2015, \mnras, 454, 4282,
  \dodoi{10.1093/mnras/stv2248}

\bibitem[{{Marsh} {et~al.}(2017){Marsh}, {Whitworth}, {Lomax}, {Ragan},
  {Becciani}, {Cambr{\'e}sy}, {Di Giorgio}, {Eden}, {Elia}, \&
  {Kacsuk}}]{2017MNRAS.471.2730M}
{Marsh}, K.~A., {Whitworth}, A.~P., {Lomax}, O., {et~al.} 2017, \mnras, 471,
  2730, \dodoi{10.1093/mnras/stx1723}

\bibitem[{{Massey} \& {Thompson}(1991)}]{1991AJ....101.1408M}
{Massey}, P., \& {Thompson}, A.~B. 1991, \aj, 101, 1408, \dodoi{10.1086/115774}

\bibitem[{{Mathieu}(1985)}]{1985IAUS..113..427M}
{Mathieu}, R.~D. 1985, in IAU Symposium, Vol. 113, Dynamics of Star Clusters,
  ed. J.~{Goodman} \& P.~{Hut}, 427--446

\bibitem[{{Matsakis} {et~al.}(1976){Matsakis}, {Evans}, {Sato}, \&
  {Zuckerman}}]{1976AJ.....81..172M}
{Matsakis}, D.~N., {Evans}, N.~J., I., {Sato}, T., \& {Zuckerman}, B. 1976,
  \aj, 81, 172, \dodoi{10.1086/111871}

\bibitem[{{McNamara} \& {Sekiguchi}(1986)}]{1986ApJ...310..613M}
{McNamara}, B.~J., \& {Sekiguchi}, K. 1986, \apj, 310, 613,
  \dodoi{10.1086/164714}

\bibitem[{{Melikian} \& {Shevchenko}(1990)}]{1990Afz....32..169M}
{Melikian}, N.~D., \& {Shevchenko}, V.~S. 1990, Astrofizika, 32, 169

\bibitem[{{Mermilliod}(2000)}]{2000ASPC..211...43M}
{Mermilliod}, J.-C. 2000, in Astronomical Society of the Pacific Conference
  Series, Vol. 211, Massive Stellar Clusters, ed. A.~{Lan{\c c}on} \& C.~M.
  {Boily}, 43

\bibitem[{{Molinari} {et~al.}(2010){Molinari}, {Swinyard}, {Bally}, {Barlow},
  {Bernard}, {Martin}, {Moore}, {Noriega-Crespo}, {Plume}, \&
  {Testi}}]{2010PASP..122..314M}
{Molinari}, S., {Swinyard}, B., {Bally}, J., {et~al.} 2010, \pasp, 122, 314,
  \dodoi{10.1086/651314}

\bibitem[{{Panagia}(1973)}]{1973AJ.....78..929P}
{Panagia}, N. 1973, \aj, 78, 929, \dodoi{10.1086/111498}

\bibitem[{{Pandey} {et~al.}(2001){Pandey}, {Nilakshi}, {Ogura}, {Sagar}, \&
  {Tarusawa}}]{2001A&A...374..504P}
{Pandey}, A.~K., {Nilakshi}, {Ogura}, K., {Sagar}, R., \& {Tarusawa}, K. 2001,
  \aap, 374, 504, \dodoi{10.1051/0004-6361:20010642}

\bibitem[{{Pandey} {et~al.}(2000){Pandey}, {Ogura}, \&
  {Sekiguchi}}]{2000PASJ...52..847P}
{Pandey}, A.~K., {Ogura}, K., \& {Sekiguchi}, K. 2000, \pasj, 52, 847,
  \dodoi{10.1093/pasj/52.5.847}

\bibitem[{{Pandey} {et~al.}(2020{\natexlab{a}}){Pandey}, {Sharma}, {Kobayashi},
  {Sarugaku}, \& {Ogura}}]{2020MNRAS.492.2446P}
{Pandey}, A.~K., {Sharma}, S., {Kobayashi}, N., {Sarugaku}, Y., \& {Ogura}, K.
  2020{\natexlab{a}}, \mnras, 492, 2446, \dodoi{10.1093/mnras/stz3596}

\bibitem[{{Pandey} {et~al.}(2008){Pandey}, {Sharma}, {Ogura}, {Ojha}, {Chen},
  {Bhatt}, \& {Ghosh}}]{2008MNRAS.383.1241P}
{Pandey}, A.~K., {Sharma}, S., {Ogura}, K., {et~al.} 2008, \mnras, 383, 1241,
  \dodoi{10.1111/j.1365-2966.2007.12641.x}

\bibitem[{{Pandey} {et~al.}(2003){Pandey}, {Upadhyay}, {Nakada}, \&
  {Ogura}}]{2003AA...397..191P}
{Pandey}, A.~K., {Upadhyay}, K., {Nakada}, Y., \& {Ogura}, K. 2003, \aap, 397,
  191, \dodoi{10.1051/0004-6361:20021509}

\bibitem[{{Pandey} {et~al.}(2005){Pandey}, {Upadhyay}, {Ogura}, {Sagar},
  {Mohan}, {Mito}, {Bhatt}, \& {Bhatt}}]{2005MNRAS.358.1290P}
{Pandey}, A.~K., {Upadhyay}, K., {Ogura}, K., {et~al.} 2005, \mnras, 358, 1290,
  \dodoi{10.1111/j.1365-2966.2005.08784.x}

\bibitem[{{Pandey} {et~al.}(2013){Pandey}, {Eswaraiah}, {Sharma}, {Samal},
  {Chauhan}, {Chen}, {Jose}, {Ojha}, {Kesh Yadav}, \&
  {Chandola}}]{2013ApJ...764..172P}
{Pandey}, A.~K., {Eswaraiah}, C., {Sharma}, S., {et~al.} 2013, \apj, 764, 172,
  \dodoi{10.1088/0004-637X/764/2/172}

\bibitem[{{Pandey} {et~al.}(2020{\natexlab{b}}){Pandey}, {Sharma}, {Panwar},
  {Dewangan}, {Ojha}, {Bisen}, {Sinha}, {Ghosh}, \&
  {Pandey}}]{2020ApJ...891...81P}
{Pandey}, R., {Sharma}, S., {Panwar}, N., {et~al.} 2020{\natexlab{b}}, \apj,
  891, 81, \dodoi{10.3847/1538-4357/ab6dc7}

\bibitem[{{Panwar} {et~al.}(2018){Panwar}, {Pandey}, {Samal}, {Battinelli},
  {Ogura}, {Ojha}, {Chen}, \& {Singh}}]{2018AJ....155...44P}
{Panwar}, N., {Pandey}, A.~K., {Samal}, M.~R., {et~al.} 2018, \aj, 155, 44,
  \dodoi{10.3847/1538-3881/aa9f1b}

\bibitem[{{Pastorelli} {et~al.}(2019){Pastorelli}, {Marigo}, {Girardi}, {Chen},
  {Rubele}, {Trabucchi}, {Aringer}, {Bladh}, {Bressan}, {Montalb{\'a}n},
  {Boyer}, {Dalcanton}, {Eriksson}, {Groenewegen}, {H{\"o}fner}, {Lebzelter},
  {Nanni}, {Rosenfield}, {Wood}, \& {Cioni}}]{2019MNRAS.485.5666P}
{Pastorelli}, G., {Marigo}, P., {Girardi}, L., {et~al.} 2019, \mnras, 485,
  5666, \dodoi{10.1093/mnras/stz725}

\bibitem[{{Pecaut} \& {Mamajek}(2013)}]{2013ApJS..208....9P}
{Pecaut}, M.~J., \& {Mamajek}, E.~E. 2013, \apjs, 208, 9,
  \dodoi{10.1088/0067-0049/208/1/9}

\bibitem[{{Perren} {et~al.}(2015){Perren}, {V{\'a}zquez}, \&
  {Piatti}}]{2015A&A...576A...6P}
{Perren}, G.~I., {V{\'a}zquez}, R.~A., \& {Piatti}, A.~E. 2015, \aap, 576, A6,
  \dodoi{10.1051/0004-6361/201424946}

\bibitem[{{Phelps} \& {Janes}(1994)}]{1994ApJS...90...31P}
{Phelps}, R.~L., \& {Janes}, K.~A. 1994, \apjs, 90, 31, \dodoi{10.1086/191857}

\bibitem[{{Piano} {et~al.}(2019){Piano}, {Cardillo}, {Pilia}, {Trois},
  {Giuliani}, {Bulgarelli}, {Parmiggiani}, \& {Tavani}}]{2019ApJ...878...54P}
{Piano}, G., {Cardillo}, M., {Pilia}, M., {et~al.} 2019, \apj, 878, 54,
  \dodoi{10.3847/1538-4357/ab1f69}

\bibitem[{{Planck Collaboration} {et~al.}(2014){Planck Collaboration}, {Ade},
  {Aghanim}, {Armitage-Caplan}, {Arnaud}, {Ashdown}, {Atrio-Barand ela},
  {Aumont}, {Baccigalupi}, \& {Banday}}]{planck1234}
{Planck Collaboration}, {Ade}, P.~A.~R., {Aghanim}, N., {et~al.} 2014, \aap,
  571, A9, \dodoi{10.1051/0004-6361/201321531}

\bibitem[{{Reddish} {et~al.}(1966){Reddish}, {Lawrence}, \&
  {Pratt}}]{1966PROE....5..111R}
{Reddish}, V.~C., {Lawrence}, L.~C., \& {Pratt}, N.~M. 1966, Publications of
  the Royal Observatory of Edinburgh, 5, 111

\bibitem[{{Reipurth} \& {Schneider}(2008)}]{2008hsf1.book...36R}
{Reipurth}, B., \& {Schneider}, N. 2008, {Star Formation and Young Clusters in
  Cygnus}, ed. B.~{Reipurth}, 36

\bibitem[{{Sagar} \& {Richtler}(1991)}]{1991A&A...250..324S}
{Sagar}, R., \& {Richtler}, T. 1991, \aap, 250, 324

\bibitem[{{Salpeter}(1955)}]{1955ApJ...121..161S}
{Salpeter}, E.~E. 1955, \apj, 121, 161, \dodoi{10.1086/145971}

\bibitem[{{Samal} {et~al.}(2007){Samal}, {Pandey}, {Ojha}, {Ghosh}, {Kulkarni},
  \& {Bhatt}}]{2007ApJ...671..555S}
{Samal}, M.~R., {Pandey}, A.~K., {Ojha}, D.~K., {et~al.} 2007, \apj, 671, 555,
  \dodoi{10.1086/522941}

\bibitem[{{Sandell} {et~al.}(2012){Sandell}, {Wiesemeyer}, {Requena-Torres},
  {Heyminck}, {G{\"u}sten}, {Stutzki}, {Simon}, \&
  {Graf}}]{2012A&A...542L..14S}
{Sandell}, G., {Wiesemeyer}, H., {Requena-Torres}, M.~A., {et~al.} 2012, \aap,
  542, L14, \dodoi{10.1051/0004-6361/201218920}

\bibitem[{{Sariya} {et~al.}(2017){Sariya}, {Jiang}, \&
  {Yadav}}]{2017AJ....153..134S}
{Sariya}, D.~P., {Jiang}, I.-G., \& {Yadav}, R.~K.~S. 2017, \aj, 153, 134,
  \dodoi{10.3847/1538-3881/aa5be6}

\bibitem[{{Sariya} \& {Yadav}(2015)}]{2015A&A...584A..59S}
{Sariya}, D.~P., \& {Yadav}, R.~K.~S. 2015, \aap, 584, A59,
  \dodoi{10.1051/0004-6361/201526688}

\bibitem[{{Sariya} {et~al.}(2012){Sariya}, {Yadav}, \&
  {Bellini}}]{2012A&A...543A..87S}
{Sariya}, D.~P., {Yadav}, R.~K.~S., \& {Bellini}, A. 2012, \aap, 543, A87,
  \dodoi{10.1051/0004-6361/201219306}

\bibitem[{{Scalo}(1998)}]{1998ASPC..142..201S}
{Scalo}, J. 1998, in Astronomical Society of the Pacific Conference Series,
  Vol. 142, The Stellar Initial Mass Function (38th Herstmonceux Conference),
  ed. G.~{Gilmore} \& D.~{Howell}, 201

\bibitem[{{Scalo}(1986)}]{1986FCPh...11....1S}
{Scalo}, J.~M. 1986, \fcp, 11, 1

\bibitem[{{Schneider} {et~al.}(2006){Schneider}, {Bontemps}, {Simon}, {Jakob},
  {Motte}, {Miller}, {Kramer}, \& {Stutzki}}]{2006A&A...458..855S}
{Schneider}, N., {Bontemps}, S., {Simon}, R., {et~al.} 2006, \aap, 458, 855,
  \dodoi{10.1051/0004-6361:20065088}

\bibitem[{{Schneider} {et~al.}(2007){Schneider}, {Simon}, {Bontemps},
  {Comer{\'o}n}, \& {Motte}}]{2007A&A...474..873S}
{Schneider}, N., {Simon}, R., {Bontemps}, S., {Comer{\'o}n}, F., \& {Motte}, F.
  2007, \aap, 474, 873, \dodoi{10.1051/0004-6361:20077540}

\bibitem[{{Sharma} {et~al.}(2008){Sharma}, {Pandey}, {Ogura}, {Aoki}, {Pandey},
  {Sandhu}, \& {Sagar}}]{2008AJ....135.1934S}
{Sharma}, S., {Pandey}, A.~K., {Ogura}, K., {et~al.} 2008, \aj, 135, 1934,
  \dodoi{10.1088/0004-6256/135/5/1934}

\bibitem[{{Sharma} {et~al.}(2006){Sharma}, {Pandey}, {Ogura}, {Mito},
  {Tarusawa}, \& {Sagar}}]{2006AJ....132.1669S}
---. 2006, \aj, 132, 1669, \dodoi{10.1086/507094}

\bibitem[{{Sharma} {et~al.}(2017){Sharma}, {Pandey}, {Ojha}, {Bhatt}, {Ogura},
  {Kobayashi}, {Yadav}, \& {Pandey}}]{2017MNRAS.467.2943S}
{Sharma}, S., {Pandey}, A.~K., {Ojha}, D.~K., {et~al.} 2017, \mnras, 467, 2943,
  \dodoi{10.1093/mnras/stx014}

\bibitem[{{Sharma} {et~al.}(2007){Sharma}, {Pandey}, {Ojha}, {Chen}, {Ghosh},
  {Bhatt}, {Maheswar}, \& {Sagar}}]{2007MNRAS.380.1141S}
---. 2007, \mnras, 380, 1141, \dodoi{10.1111/j.1365-2966.2007.12156.x}

\bibitem[{{Sharma} {et~al.}(2012){Sharma}, {Pandey}, {Pandey}, {Chauhan},
  {Ogura}, {Ojha}, {Borrissova}, {Mito}, {Verdugo}, \&
  {Bhatt}}]{2012PASJ...64..107S}
{Sharma}, S., {Pandey}, A.~K., {Pandey}, J.~C., {et~al.} 2012, \pasj, 64, 107,
  \dodoi{10.1093/pasj/64.5.107}

\bibitem[{{Shevchenko} {et~al.}(1991){Shevchenko}, {Ibragimov}, \&
  {Chenysheva}}]{1991SvA....35..229S}
{Shevchenko}, V.~S., {Ibragimov}, M.~A., \& {Chenysheva}, T.~L. 1991, \sovast,
  35, 229

\bibitem[{{Siess} {et~al.}(2000){Siess}, {Dufour}, \&
  {Forestini}}]{2000AA...358..593S}
{Siess}, L., {Dufour}, E., \& {Forestini}, M. 2000, \aap, 358, 593

\bibitem[{{Skrutskie} {et~al.}(2006){Skrutskie}, {Cutri}, {Stiening},
  {Weinberg}, {Schneider}, {Carpenter}, {Beichman}, {Capps}, {Chester},
  {Elias}, {Huchra}, {Liebert}, {Lonsdale}, {Monet}, {Price}, {Seitzer},
  {Jarrett}, {Kirkpatrick}, {Gizis}, {Howard}, {Evans}, {Fowler}, {Fullmer},
  {Hurt}, {Light}, {Kopan}, {Marsh}, {McCallon}, {Tam}, {Van Dyk}, \&
  {Wheelock}}]{2006AJ....131.1163S}
{Skrutskie}, M.~F., {Cutri}, R.~M., {Stiening}, R., {et~al.} 2006, \aj, 131,
  1163, \dodoi{10.1086/498708}

\bibitem[{{Spitzer} \& {Hart}(1971)}]{1971ApJ...164..399S}
{Spitzer}, Jr., L., \& {Hart}, M.~H. 1971, \apj, 164, 399,
  \dodoi{10.1086/150855}

\bibitem[{{Stetson}(1987)}]{1987PASP...99..191S}
{Stetson}, P.~B. 1987, \pasp, 99, 191, \dodoi{10.1086/131977}

\bibitem[{{Stetson}(1990)}]{1990PASP..102..932S}
---. 1990, \pasp, 102, 932, \dodoi{10.1086/132719}

\bibitem[{{Stetson}(1992)}]{1992ASPC...25..297S}
{Stetson}, P.~B. 1992, in Astronomical Society of the Pacific Conference
  Series, Vol.~25, Astronomical Data Analysis Software and Systems I, ed. D.~M.
  {Worrall}, C.~{Biemesderfer}, \& J.~{Barnes}, 297

\bibitem[{{Tan} {et~al.}(2014){Tan}, {Beltr{\'a}n}, {Caselli}, {Fontani},
  {Fuente}, {Krumholz}, {McKee}, \& {Stolte}}]{2014prpl.conf..149T}
{Tan}, J.~C., {Beltr{\'a}n}, M.~T., {Caselli}, P., {et~al.} 2014, in Protostars
  and Planets VI, ed. H.~{Beuther}, R.~S. {Klessen}, C.~P. {Dullemond}, \&
  T.~{Henning}, 149

\bibitem[{{Taylor} {et~al.}(2003){Taylor}, {Gibson}, {Peracaula}, {Martin},
  {Landecker}, {Brunt}, {Dewdney}, {Dougherty}, {Gray}, {Higgs}, {Kerton},
  {Knee}, {Kothes}, {Purton}, {Uyaniker}, {Wallace}, {Willis}, \&
  {Durand}}]{2003AJ....125.3145T}
{Taylor}, A.~R., {Gibson}, S.~J., {Peracaula}, M., {et~al.} 2003, \aj, 125,
  3145, \dodoi{10.1086/375301}

\bibitem[{{Thompson} {et~al.}(2012){Thompson}, {Urquhart}, {Moore}, \&
  {Morgan}}]{2012MNRAS.421..408T}
{Thompson}, M.~A., {Urquhart}, J.~S., {Moore}, T.~J.~T., \& {Morgan}, L.~K.
  2012, \mnras, 421, 408, \dodoi{10.1111/j.1365-2966.2011.20315.x}

\bibitem[{{Tibaldo} \& {Grenier}(2013)}]{2013NuPhS.239...70T}
{Tibaldo}, L., \& {Grenier}, I.~A. 2013, Nuclear Physics B Proceedings
  Supplements, 239, 70, \dodoi{10.1016/j.nuclphysbps.2013.05.011}

\bibitem[{{Tody}(1986)}]{1986SPIE..627..733T}
{Tody}, D. 1986, Society of Photo-Optical Instrumentation Engineers (SPIE)
  Conference Series, Vol. 627, {The IRAF Data Reduction and Analysis System},
  ed. D.~L. {Crawford}, 733

\bibitem[{{Tody}(1993)}]{1993ASPC...52..173T}
---. 1993, Astronomical Society of the Pacific Conference Series, Vol.~52,
  {IRAF in the Nineties}, ed. R.~J. {Hanisch}, R.~J.~V. {Brissenden}, \&
  J.~{Barnes}, 173

\bibitem[{{Turner}(1976)}]{1976AJ.....81.1125T}
{Turner}, D.~G. 1976, \aj, 81, 1125, \dodoi{10.1086/111994}

\bibitem[{{Uchiyama} {et~al.}(2002){Uchiyama}, {Takahashi}, {Aharonian}, \&
  {Mattox}}]{2002ApJ...571..866U}
{Uchiyama}, Y., {Takahashi}, T., {Aharonian}, F.~A., \& {Mattox}, J.~R. 2002,
  \apj, 571, 866, \dodoi{10.1086/340121}

\bibitem[{{Vansevicius}(1992)}]{1992BaltA...1...31V}
{Vansevicius}, V. 1992, Baltic Astronomy, 1, 31,
  \dodoi{10.1515/astro-1992-0106}

\bibitem[{{Walker} \& {Hodge}(1968)}]{1968PASP...80..290W}
{Walker}, G.~A.~H., \& {Hodge}, S.~M. 1968, \pasp, 80, 290,
  \dodoi{10.1086/128631}

\bibitem[{{Watson} {et~al.}(2010){Watson}, {Hanspal}, \&
  {Mengistu}}]{2010ApJ...716.1478W}
{Watson}, C., {Hanspal}, U., \& {Mengistu}, A. 2010, \apj, 716, 1478,
  \dodoi{10.1088/0004-637X/716/2/1478}

\bibitem[{{Wendker} {et~al.}(1983){Wendker}, {Schramm}, \&
  {Dieckvoss}}]{1983A&A...121...69W}
{Wendker}, H.~J., {Schramm}, K.~J., \& {Dieckvoss}, C. 1983, \aap, 121, 69

\bibitem[{{Whittet}(2003)}]{2003dge..conf.....W}
{Whittet}, D.~C.~B., ed. 2003, {Dust in the galactic environment}

\bibitem[{{Whitworth} {et~al.}(1994){Whitworth}, {Bhattal}, {Chapman},
  {Disney}, \& {Turner}}]{1994MNRAS.268..291W}
{Whitworth}, A.~P., {Bhattal}, A.~S., {Chapman}, S.~J., {Disney}, M.~J., \&
  {Turner}, J.~A. 1994, \mnras, 268, 291, \dodoi{10.1093/mnras/268.1.291}

\bibitem[{{Wright} {et~al.}(2010){Wright}, {Eisenhardt}, {Mainzer}, {Ressler},
  {Cutri}, {Jarrett}, {Kirkpatrick}, {Padgett}, {McMillan}, {Skrutskie},
  {Stanford}, {Cohen}, {Walker}, {Mather}, {Leisawitz}, {Gautier}, {McLean},
  {Benford}, {Lonsdale}, {Blain}, {Mendez}, {Irace}, {Duval}, {Liu}, {Royer},
  {Heinrichsen}, {Howard}, {Shannon}, {Kendall}, {Walsh}, {Larsen}, {Cardon},
  {Schick}, {Schwalm}, {Abid}, {Fabinsky}, {Naes}, \&
  {Tsai}}]{2010AJ....140.1868W}
{Wright}, E.~L., {Eisenhardt}, P.~R.~M., {Mainzer}, A.~K., {et~al.} 2010, \aj,
  140, 1868, \dodoi{10.1088/0004-6256/140/6/1868}

\bibitem[{{Yadav} {et~al.}(2013){Yadav}, {Sariya}, \&
  {Sagar}}]{2013MNRAS.430.3350Y}
{Yadav}, R.~K.~S., {Sariya}, D.~P., \& {Sagar}, R. 2013, \mnras, 430, 3350,
  \dodoi{10.1093/mnras/stt136}

\bibitem[{{Zinnecker} \& {Yorke}(2007)}]{2007ARA&A..45..481Z}
{Zinnecker}, H., \& {Yorke}, H.~W. 2007, \araa, 45, 481,
  \dodoi{10.1146/annurev.astro.44.051905.092549}

\end{thebibliography}
	\bibliographystyle{aasjournal}


	\begin{table*}
	\scriptsize
	\centering
	\caption{\label{log}  Log of optical observations with the 104-cm Sampurnanand telescope, Nainital.}
	\begin{tabular}{@{}rr@{}}
	\hline
	Date of observations/Filter& Exp. (sec)$\times$ No. of frames\\
	\hline
	&SA98\\
	04 November 2005\\
	$U$   &  $180\times1,300\times9$\\
	$B$   &  $120\times3,180\times8$\\
	$V$   &  $60\times8,120\times3$\\
	$R_c$ &  $30\times9,60\times3$\\
	$I_c$ &  $30\times10,60\times2$\\
	\\
	&NGC 6910(Center)\\
	$U$   &  $60\times3,300\times3$\\
	$B$   &  $30\times3,180\times3$\\
	$V$   &  $60\times3,120\times3$\\
	$R_c$ &  $10\times3,60\times3$\\
	$I_c$ &  $10\times3,60\times3$\\
	\\
	&NGC 6910(Center)\\
	13 June 2005\\
	$U$   &  $300\times3,1200\times3$\\
	$B$   &  $300\times3,900\times3$\\
	$V$   &  $10\times3,60\times3,900\times3$\\
	$R_c$ &  $10\times1,300\times4$\\
	$I_c$ &  $0\times1,1\times1,2\times1,5\times3$\\
	\\
	&NGC 6910(F1)\\
	28 September 2006\\
	$U$   &  $300\times4,1200\times3$\\
	$B$   &  $60\times3,900\times3$\\
	$V$   &  $10\times1,60\times3,900\times3$\\
	$I_c$ &  $10\times3,300\times4$\\
	\\
	&NGC 6910(F2)\\
	29 September 2006\\
	$U$   &  $300\times2,1200\times2$\\
	$B$   &  $60\times3,900\times2$\\
	$V$   &  $60\times3,900\times2$\\
	$I_c$ &  $10\times2,60\times3,300\times4$\\
	\\
	&NGC 6910(F3)\\
	15 June 2005\\
	$U$   &  $60\times1,300\times3,1200\times3$\\
	$B$   &  $10\times3,60\times3,900\times3$\\
	$V$   &  $30\times3,60\times1,900\times3$\\
	$I_c$ &  $10\times3,20\times1,60\times1,300\times4$\\
	\\
	&NGC 6910(F4)\\
	26 September 2006\\
	$U$   &  $300\times3$\\
	$B$   &  $120\times4$\\
	$V$   &  $120\times3$\\
	$I_c$ &  $10\times3,30\times1,300\times4$\\
	\\
	&NGC 6910(F4)\\
	27 September 2006\\
	$U$   &  $300\times3,1200\times3$\\
	$B$   &  $60\times3,900\times3$\\
	$V$   &  $60\times3,900\times3$\\

	\hline
	\end{tabular}
	\end{table*}

	\begin{table*}
	\centering
	\scriptsize
	\caption{\label{surveys} List of surveys adopted in present work (NIR to radio wavelength).}
	\begin{tabular}{@{}l@{}c@{}c@{}c@{}}
	\hline
	Survey        &  wavelength(s)          &Resolution           & References    \\
	\hline

	Two Micron All Sky Survey (2MASS)                   &  1.25-2.2 $\mu$m               &      $\sim$2$^\prime$$^\prime$.5                                  &   \citet{2006AJ....131.1163S}    \\
	UKIRT NIR Galactic Plane Survey (GPS)               &  1.25-2.22 $\mu$m             &      $\sim$0$^\prime$$^\prime$.8                                  &  \citet{2007MNRAS.379.1599L}     \\
	{\it Spitzer} Spitzer Enhanced Imaging Products (SEIP)  &   3.6, 4.5, 5.8 and 8 $\mu$m  &  $\sim$2$^\prime$$^\prime$, $\sim$2$^\prime$$^\prime$, $\sim$2$^\prime$$^\prime$, $\sim$2$^\prime$$^\prime$          &   $^a$ \\
	{\it Wide-field Infrared Survey Explorer} (WISE)    &  3.4, 4.6, 12 and 22 $\mu$m    &      $\sim$6$^\prime$$^\prime$.1, $\sim$6$^\prime$$^\prime$.4, $\sim$6$^\prime$$^\prime$.5, $\sim$12$^\prime$$^\prime$  & \citet{2010AJ....140.1868W} \\
	{\it Spitzer} MIPS Inner Galactic Plane Survey (MIPSGAL) &  24 $\mu$m                &     $\sim$6$^\prime$$^\prime$                                     &  \citet{2005AAS...207.6333C}  \\
	{\it Herschel} Infrared Galactic Plane Survey (Hi-GAL) & 70, 160, 250, 350, 500 $\mu$m &   $\sim$5$^\prime$$^\prime$.8, $\sim$12$^\prime$$^\prime$, $\sim$18$^\prime$$^\prime$, $\sim$25$^\prime$$^\prime$, $\sim$37$^\prime$$^\prime$  & \citet{2010PASP..122..314M}  \\
	{\it Planck} polarization data                         & 850 $\mu$m                     &  $\sim$294$^\prime$$^\prime$     & \citet{planck1234}  \\
	CO survey Archive                                   &  2.6\ mm                        &     $\sim$8$^\prime$.4             &   \citet{2001ApJ...547..792D}    \\
	NRAO VLA Sky Survey (NVSS)                          &    21\ cm                      &     $\sim$46$^\prime$$^\prime$                &  \citet{1998AJ....115.1693C}   \\
	Canadian Galactic Plane Survey (CGPS)               &  21\ cm, 74\ cm                  &     $1^\prime\times1^\prime$ $csc\delta$, $3^\prime.4\times3^\prime.4$ $csc\delta $                                          &  \citet{2003AJ....125.3145T}  \\
	GAIA DR2 (magnitudes, Parallax and Proper motion)   & 330-1050nm                      &      0.4 mas   & \citet{gaia1234,gaia5678}      \\
	\hline
	\end{tabular}

	$^a$: https://irsa.ipac.caltech.edu/data/SPITZER/Enhanced/SEIP/overview.html 
	\end{table*}

\begin{table}[]
\centering
\scriptsize
\caption{Sample of the 128 cluster members identified through their $Gaia$ proper motion data. Complete table is available in the electronic form only.}
\label{PMT}
	\begin{tabular}{@{}l@{ }c@{ }c@{ }c@{ }c@{ }c@{ }c@{ }c@{ }c@{ }c@{ }c@{ }c@{ }c@{ }c@{}}
\hline
	ID& $\alpha_{J2000}$  &  $\delta_{J2000}$   & $V$      & $B$    & $I$   &  $R$    & $U$  &  Parallax   & $\mu_\alpha$cos($\delta$)    &  $\mu_\delta$ & $G$ & $(G_{BP} -G_{RP})$ & P$_\mu$  \\
	&(deg) & (deg) & (mag) & (mag) & (mag) & (mag) & (mag) &  (mas) & (mas yr$^{-1}$) & (mas yr$^{-1}$) & (mag) & (mag) & (\%)  \\
\hline
	1 & 305.888133  &  40.693244   &  18.376$\pm$0.011   &   19.838$\pm$0.015  & 16.211$\pm$0.010  &   -              &         -          &    0.753 $\pm$0.131  &  -2.706$\pm$0.148  &  -5.341$\pm$0.197  &   17.450   &  2.080  &    98     \\
	2 & 305.898745  &  40.707285   &  19.526$\pm$0.014   &   21.124$\pm$0.033  & 17.243$\pm$0.010  &   -              &         -          &    0.374 $\pm$0.216  &  -2.408$\pm$0.225  &  -5.536$\pm$0.323  &   18.504   &  2.145  &    96     \\
	3 & 305.875944  &  40.712293   &  19.277$\pm$0.013   &   20.818$\pm$0.023  & 16.937$\pm$0.013  &   -              &         -          &    0.453 $\pm$0.161  &  -2.974$\pm$0.183  &  -5.727$\pm$0.247  &   18.245   &  2.167  &    81     \\
	4 & 305.797771  &  40.716103   &  19.653$\pm$0.013   &   21.479$\pm$0.052  & 17.105$\pm$0.060  & 18.307$\pm$0.027 &         -          &    0.389 $\pm$0.205  &  -2.527$\pm$0.306  &  -5.285$\pm$0.314  &   18.427   &  2.377  &    96     \\
\hline
\end{tabular}
\end{table}

	\begin{table}[]
	\centering
	\caption{A summary of the physical parameters of the cluster.}
	\label{ST}
	\begin{tabular}{@{}l@{}c@{}c@{}c@{}c@{}c@{}}
	\hline
	Reference &  $R_{cl}$  & Age &  Distance modulus & $R_{V}$ &  $E(B-V)$   \\
       &  (arcmin) & (Myr) &  (mag)&  & (mag)   \\
       \hline
       Present study $(UBVRI)$ & 5.5 & 4.5$\pm$2.5 & 11.18$\pm$0.12 & 3.75$\pm$0.02 & 0.95$\pm$0.10 \\
       \citet[][$UBV$]{2000AJ....119.1848D} & $-$ & 6.5$\pm$3.0   & 11.2$\pm$ 0.2 &   3.1$\pm$0.1     & 1.02$\pm$ 0.13   \\
       \citet[][$VI$]{kola1234} & $-$  & 6.0$\pm$2.0  & 11.0$\pm$0.3  & $-$   &  1.19$\pm$0.21  \\
       \hline
	\end{tabular}
	\end{table}

	\begin{table*}
	\centering
	\caption{\label{mfslope}The mass function slope for two sub-regions and for the
	whole cluster region in the given mass range.}
	\begin{tabular}{@{}cccccccc@{}}
	\hline
		Mass range &  \multicolumn{3}{c}{Mass Function slopes ($\Gamma$)} & \\
	 ($M_\odot$) & Core region & Corona region & Cluster region & \\
	\hline
	 $24.86-0.8$ & $-0.42\pm0.20$  & $-0.69\pm0.14$ & $-0.74\pm0.15$ & \\
	 $24.86-2.3$ & $-0.38\pm0.39$  & $-0.54\pm0.28$ & $-0.58\pm0.25$ & \\
	\hline
	\end{tabular}
	\end{table*}






%
%

\end{document}